\documentclass[onecolumn]{aa}  
\usepackage{graphicx}
\usepackage{txfonts}
\usepackage{psfig}
\newfont{\myfont}{cmmib10}

\newcommand{\balpha}{\hbox{\myfont \symbol{11} }}
\newcommand{\bbeta}{\hbox{\myfont \symbol{12} }}

\newcommand{\bxi}{\hbox{\myfont \symbol{24} }}
\newcommand{\brho}{\hbox{\myfont \symbol{26} }}
\newfont{\myfontsmall}{cmmib8}

\newcommand{\brhosm}{\hbox{\myfontsmall \symbol{26} }}

\DeclareSymbolFont{cmmi}{OML}{cmm}{m}{it}
\DeclareMathSymbol{v}{\mathalpha}{cmmi}{"76}

\begin{document}
   \title{Scattering of Gravitational Radiation}

   \subtitle{Intensity Fluctuations}

   \author{Jean-Pierre Macquart
          \thanks{NRAO Jansky Fellow}
          }

   \offprints{J.-P. Macquart}

   \institute{National Radio Astronomy Observatory, Socorro NM 87801, U.S.A. \\
                   Department of Astronomy, California Institute of Technology, MS 105-24, Pasadena, CA 				91125, U.S.A. \\
                   \email{jpm@astro.caltech.edu}
             }

   \date{}

 
  \abstract
   {}
   {The effect of gravitational microlensing on the intensity of gravitational radiation as it propagates through an inhomogeneous medium is considered. Lensing by both stars and a power law spectrum of density perturbations is examined.}
   {The long wavelengths characteristic of gravitational radiation mandate a statistical, physical-optics approach to treat the effect of the lensing.
   }
   {A model for the mass power spectrum of a starfield, including the effects of clustering and allowing for a distribution of stellar masses, is constructed and used to determine both the amplitude of fluctuations in the gravitational wave strain and its associated temporal fluctuation spectrum.  For a uniformly distributed starfield the intensity variance scales linearly with stellar density, $\sigma$, but is enhanced by a factor $\ga \sigma r_{\rm F}^2$ when clustering is important, where $r_{\rm F}$ is the Fresnel scale.  The effect of lensing by a power law mass spectrum, applicable to lensing by small scale fluctuations in gas and dark matter, is also considered.
 For power law mass density spectra with indices steeper than $-2$ the wave amplitude exhibits rms fluctuations  $1.3 \langle \Delta \Sigma^2\rangle^{1/4} (D_{\rm eff}/1\,{\rm Gpc})^{1/2} $\%, where $\langle \Delta \Sigma^2 \rangle$ is the variance in the mass surface density measured in $M_\odot^2 \,{\rm pc}^{-4}$ and $D_{\rm eff}$ is the effective distance to the lensing medium.  For shallower spectra the amplitude of the fluctuations depends additionally on the inner length scale and power law index of the density fluctuations.  The intensity fluctuations are dominated by temporal fluctuations on long timescales.  For lensing material moving at a speed $v$ across the line of sight the fluctuation timescale exceeds $v^{-1} (D_{\rm eff} \lambda)^{1/2}$. Lensing by small scale structure induces at most $\approx 15\,$\% rms variations if the line of sight to a gravitational wave source intersects a region with densities $\sim 100 M_\odot {\rm pc}^{-2}$, which are typically encountered in the vicinity of galaxy clusters.}
{}  


   \maketitle
%

\section{Introduction}

The amplification of electromagnetic radiation due to gravitational microlensing (Byalko 1969; Chang \& Refsdal 1979, 1984; Paczy\'nski 1986) is now a well-established phenomenon that has proven to be a successful tool in probing planetary- to solar-mass objects in the vicinity of the Milky Way (e.g. Afonso et al. 2000).   Gravitational radiation is also subject to lensing by small-scale structure, but such lensing differs from its electromagnetic analogue in several qualitative respects.  In particular, under most circumstances it demands a treatment based on physical optics rather than geometric optics, and the number of objects contributing instantaneously to the lensed signal can greatly exceed that expected in the electromagnetic case.

The lensing of gravitational radiation occurs in a regime in which wave effects cannot be ignored (see Macquart 2004, hereafter M04, and references therein).  Gravitational wave detectors such as LIGO operate at frequencies $10 - 10^4$\,Hz, while the planned Laser Interferometer Space Antenna (LISA) will operate from $10^{-1} - 10^{-4}$\,Hz.  Thus the size of most objects likely to lens such radiation is large compared to the wavelength of the radiation.  A sufficient (but not necessary) condition for diffractive effects to be important is that the wavelength exceeds the sizes of the lensing objects, which occurs for objects of mass $\la 10^5 \, \nu^{-1}$\,M$_\odot$ (Takahashi \& Nakamura 2003). 

The radius of the region that contributes to wave fluctuations is characterised by the Fresnel scale, $r_{\rm F} \sim \sqrt{D/k}$, where $D$ is an effective distance to the lensing material and $k$ is the wavenumber (M04).  This quantity exceeds $1\,$pc for typical radiation wavelengths and lensing geometries (i.e. with $D$ characteristic of lensing at cosmological distances), so that many tens to thousands of objects (e.g. stars), depending on the wavelength, may contribute to the lensing.  It is often therefore impractical to specify the exact mass distribution at all locations and all times, and the lensing of gravitational radiation lends itself to a treatment in which the mass density fluctuations of the lensing objects are described statistically.


Most operational and proposed detectors are directly sensitive to the wave amplitude of the gravitational radiation itself, and many gravitational lensing effects relevant to these detectors are described in terms of second-order moments of the wavefield or mutual coherence functions.  These were investigated in M04 (see also Takahashi \& Nakamura 2003), where a statistical treatment was introduced to determine the effect of lensing on the properties of the gravitational radiation.  

In the present paper this formalism is extended to compute the intensity fluctuations induced by gravitational lensing by small-scale structures.   More generally, we consider the effect of lensing on the fourth-order moment of the wavefield, of which a special case describes intensity fluctuations.  There are several reasons to consider this quantity.

The first is that the wave power, measured by the intensity, is a fundamental quantity that may be affected by gravitational lensing.  Holz \& Hughes (2005) suggest that binary black-hole mergers are suitable as  ``gravitational-wave cosmological sirens'' if their electromagnetic counterparts can be identified because the frequency of the gravitational radiation and its time derivative in such systems uniquely determines the intrinsic luminosity.  Uncertainties in the identification of the electromagnetic counterpart are estimated to introduce a 1-10\% uncertainty in the distance relation (Holz \& Hughes 2005).  However, lensing-induced intensity variability further hinders the suitability of such sources as standard sirens.  Lensing potentially causes misestimation of the source intensity and, in long lived sources, may hinder identification of those variations which are intrinsic to the source itself.  The effect of lensing by large-scale structures can be accounted for by examining the line of sight for lensing objects, but it is estimated that at least half the uncertainty is from lensing by structures less than $\sim 0.6\,$Mpc in size (Kocsis et al. 2006).  This subject has already received some attention through examination of fluctuations in the wave amplitude and phase (Takahashi 2006).

The second is that the calculation of the ensemble-average mutual coherence presented in M04 is not sufficient to fully describe the effects imposed by gravitational lensing on the wave properties.  
Practical observations of gravitational wave sources extend only over a finite duration, so that the measured mutual coherence may deviate significantly from its average value.
Thus comparison of the ensemble-average mutual coherence against its observed value may be invalid.  The fourth order moment is an estimator of the amount by which the second moment instantaneously varies from its ensemble average value because, in its most general form, it may be regarded as the cross-correlation of two instantaneous mutual coherence measurements.  We note that the fourth order moment is computed in other branches of scattering physics for this precisely this reason.  This is particularly the case in scintillation physics where, even though a large ensemble of phase fluctuations may contribute instantaneously to the scattering, the instantaneous mutual coherence can exhibit large modulations around its ensemble average value (e.g. Goodman \& Narayan 1989).

A third is that lensing-induced intensity variations can potentially be used to recover the properties of the intervening lensing material.  It was shown in M04 that the ensemble-average mutual coherence is related to the auto-covariance of lensing-induced phase delays, which is in turn related to the underlying mass power spectrum by a Fourier transform.   It is pertinent to investigate whether fourth-order moments of the wave amplitude provide supplementary information on the underlying mass distribution.

The physics of gravitational wave microlensing strongly resembles that encountered in the scattering of radio emission from pulsars and compact quasars as it propagates through density inhomogeneities in the interstellar plasma (Rickett 1977).  In light of this similarity, the terms microlensing and scattering are used interchangeably throughout the text.  The main distinguishing characteristic of the present situation is that the power spectrum of lensing-induced phase fluctuations is typically steeper than that experienced in other applications, due to the manner in which the gravitational phase delays are generated from the underlying mass fluctuations.  This qualitatively alters the nature of the intensity fluctuations.

The large wavelengths typical of gravitational radiation, in the range $ 10^4\,{\rm m} \la \lambda \la 10^{17}\,$m, introduce a number of physical effects that are usually unimportant in the microlensing of electromagnetic radiation.  In particular, the coherence area over which objects (e.g. stars) may contribute to the gravitational wave amplitude received at Earth can exceed many square parsecs so that, instead of a single object dominating the lensing at any given instant, an enormous number of objects can contribute to the lensing instantaneously.  Thus we need to consider the simultaneous cumulative effect of a large number of objects on the wave amplitude.   

An important qualitative manner in which the lensing characteristics may change is through the effect of clustering.  Since the effective area that contributes to the wavefield encompasses many stars, stellar clustering becomes important in a way that is not necessarily the case in electromagnetic lensing.  If the coherence area encompasses a cluster of $N$ stars we might expect the intensity deviations will more closely resemble lensing by a macrostar of mass $NM$ whereas  the incoherent contribution of $N$ individual stars randomly distributed in space might be only $N^{1/2} M$ if clustering were unimportant.


The properties of the sources themselves also influence the nature of the gravitational lensing.  In contrast to most lensed sources of electromagnetic radiation, sources of gravitational radiation are extremely compact and radiate ``coherently'', in the sense that they may be regarded as single emitting particles.   Their compact nature is of particular importance, since it renders their radiation far more susceptible to certain lensing effects that are not relevant to large lensed sources. For large sources amplification effects associated with structure in the lensing pattern on angular scales less than the source angular diameter are heavily suppressed (Little \& Hewish 1966; Narayan 1992); in effect the source ``washes-out'' lensing structure smaller than its own angular size.

In this paper we derive the relation between the properties of the lensed gravitational radiation and the underlying mass distribution and use it to consider lensing by stars and a power law spectrum of mass density fluctuations.  The layout of this paper is as follows.  In section 2 we review the physics of gravitational wave propagation and employ it in section 3 to develop a statistical theory for the intensity fluctuations induced by lensing on a gravitational wave.  This section also considers more general fourth-order moments of the wavefield, used to describe how the mutual coherence deviates from its mean quantity.  The theory is then applied in Sect.\,4 to lensing by a power law mass power spectrum and used to develop an intuitive understanding of the effect of lensing on gravitational wave properties.  In Sect.\,5 a model for the mass surface density power spectrum of a starfield is constructed, taking into account the clustering properties of the stars.  This model is applied in Sect.\,6 to derive the characteristics of the intensity fluctuations due to lensing by stars over a distribution of masses, and whose clustering properties may vary as a function of mass.  The implications for the properties of gravitational radiation are presented in Sect.\,7.  The results are summarized in Sect.\,8.  Although the results are derived here in the context of gravitational wave lensing they are equally applicable to any situation in which the lensing field is best described using a statistical approach since, to lowest order in the gravitational wave amplitude, the propagation of both gravitational and electromagnetic radiation are described by identical wave equations.  The physical optics treatment necessitated by the long wavelengths typical of gravitational radiation is, of course, also valid in the treatment of electromagnetic radiation.


\section{Wave Propagation}
The amplitude of gravitational radiation propagates according to the following scalar wave equation (see e.g. Takahashi \& Nakamura 2003, M04),
\begin{eqnarray}
(\nabla^2 + \omega^2 ) \tilde{\phi}  = 4 \omega^2 U({\bf r}) \tilde{\phi}, \label{WaveEqun}
\end{eqnarray}
where $\omega=2 \pi \nu$, $U({\bf r}) \ll 1$ is the gravitational potential, and $\tilde{\phi}(\nu,{\bf r})$ is the temporal Fourier transform of the scalar wave amplitude.  
We consider the thin lens geometry illustrated in Fig.\,1, for which the solution of eq. (\ref{WaveEqun}) is (e.g. Schneider 1987)
\begin{eqnarray}
\tilde{\phi}(\nu,{\bf X}) =  \frac{e^{-i \pi/2} }{2 \pi r_{\rm F}^2} \int d^2 {\bf x} \exp \left[ \frac{i}{2 r_{\rm F}^2} \left( {\bf x}- {\bf X} \right)^2 +   i \psi({\bf x})  \right], 
\label{uGrav}
\end{eqnarray}
where it is assumed that the radiation incident on the lensing plane emanates from a point source of unit amplitude.  The scaled co-ordinate, ${\bf X}$, is related to the position on the observer's plane, ${\bf x}$, by ${\bf X}=D_{LS}/D_{S} {\bf x}$, where $D_{LS}$ is the angular diameter distance from the lens to the source, and $D_S$ is the angular diameter distance from the observer to the source.  The angular diameter distance from the observer to the lens is $D_L$ and $z_L$ is the associated redshift.  Throughout the text all displacements on the observer's plane are written in terms of the scaled co-ordinate ${\bf X}$.
The Fresnel scale, defined as
\begin{eqnarray}
r_{\rm F}^2 = \frac{D_L D_{LS} \lambda }{2 \pi D_S (1 + z_L)}, \label{rF}
\end{eqnarray}
is interpreted as the scale length on the lensing plane over which geometric phase delays (i.e. phase delays in the absence of mass perturbations) become important.  The phase delay driven by fluctuations in the gravitational potential is  
\begin{eqnarray}
\psi ({\bf x}) = K \int d^2{\bf x}' \, \Sigma({\bf x}') \ln \left( \frac{|{\bf x}-{\bf x}'|}{x_0}\right), \qquad \qquad 
K = -8 \pi \frac{1+z_L}{ \lambda}   \frac{G}{c^2}, \label{psidefn}
\end{eqnarray}
where $\Sigma({\bf r})$ is the mass surface density projected onto the lensing plane.  The scale factor $x_0$ is set to unity hereafter since only phase differences across the lensing plane cause effects of interest to us here, and the total phase delay is unimportant.

The thin-lens approximation commonly employed in applications of electromagnetic lensing is still valid in the context of gravitational wave lensing despite that fact that the depths of the lensing objects may be much less than one wavelength.  This is because the extremely small amplitude of the perturbation associated with a gravitational wave renders nonlinear effects associated with sharp gradients in the gravitational potential negligible (see, e.g., Thorne 1983).  Although the lensing material may, in general, be distributed along the entire line of sight from the source to the observer, the thin-lensing approximation adopted throughout this paper is also sufficient for the lensing geometries considered here.  This approach avoids the complexity introduced by the extra mathematical machinery needed to treat phase fluctuations caused by mass inhomogeneities distributed along the entire ray path.  Moreover, extended medium treatments (Tatarski 1967; Codona \& Frehlich 1987) show that in all cases of interest here it is possible to represent the scattering physics of an extended medium in terms of an equivalent thin screen (Tatarski \& Zavorotnyi 1980).

\subsection{A statistical description of the gravitational phase delay}

An investigation of wave amplitude variations due to lensing by small scale structure requires a statistical description of the mass fluctuations that drive the phase perturbations.
Consider matter distributed with volume density $\rho({\bf R})$ with corresponding power spectrum $\Phi_\rho(q_x,q_y,q_z; z)$, where the direction of propagation is parallel to the $z$-axis, and it is assumed that the outer scale of the power spectrum is small compared to the total path length.  The Markov approximation is assumed so that changes in the shape or amplitude of the power spectrum are parameterized as a function of distance along the propagation axis.  The power spectrum of the mass perturbations projected onto an equivalent thin lensing plane is,
\begin{eqnarray}
\Phi_\Sigma({\bf q}) = \int dz \, \Phi_\rho ({\bf q},q_z=0;z),  
\end{eqnarray}
where ${\bf q}$ is the reciprocal co-ordinate to position on the two-dimensional lensing plane.  We make the simplifying assumption that the properties of the matter distribution on small scales are decoupled from the large-scale (i.e. galactic-scale) matter distribution, so that the mass surface density fluctuations are assumed to be wide-sense stationary.   This approximation is valid for the small scales over which we wish to characterize the mass power spectrum.  As such we explicitly ignore large-scale variations in the phase delay due to macrolensing by large objects, such as galaxies or clusters.  Their contribution is not relevant to the physics considered here and their effect is easily introduced separately if necessary.

The phase statistics are described by the phase autocovariance function, which quantizes the correlation of the phase between two points separated by a displacement ${\bf r}$ on the lensing plane.  This correlation is related to the underlying power spectrum of mass surface density fluctuations by,
\begin{eqnarray}
C_\psi ({\bf r}) = \langle (\psi({\bf r}'+{\bf r})- \bar \psi)(\psi({\bf r}') -\bar \psi) \rangle = K^2 \int d^2{\bf q} \, q^{-4} e^{- i {\bf q} \cdot {\bf r}} \, \Phi_\Sigma ({\bf q}). \label{CtoPhi}
\end{eqnarray}

An equivalent but often more convenient quantity is the phase structure function, which measures the mean square phase difference between two points,
\begin{eqnarray}
D_\psi ({\bf r}) = \langle [\psi({\bf r}'+{\bf r})- \psi({\bf r}')]^2 \rangle = 2 K^2 \int d^2{\bf q} \, q^{-4} \left[ 1 - e^{- i {\bf q} \cdot {\bf r}} \right] \, \Phi_\Sigma ({\bf q}). \label{DtoPhi}
\end{eqnarray}
There is a simple relationship between the phase structure function and the mean correlation between the wave amplitude measured between two receivers separated by a displacement ${\bf r}$ when the phase fluctuations obey Gaussian statistics (M04): 
\begin{eqnarray}
\langle V({\bf r}) \rangle = \exp \left[ - \frac{D_\psi \left( \frac{D_{LS}}{D_S} {\bf r} \right)}{2} \right], \label{VisEq}
\end{eqnarray}
where the incident radiation is again assumed to emanate from a point source of unit intensity.  Eq. (\ref{VisEq}) shows that the mean intensity of the gravitational radiation, obtained by setting ${\bf r}=0$, is equal to the intrinsic source intensity.  It also demonstrates that the average correlation between the wave amplitudes measured a vector ${\bf r}_2 \neq 0$ apart is reduced by the phase and amplitude fluctuations induced by lensing.  The value of this decorrelation contains information about the medium through which the radiation has propagated, and can potentially even be used to extract the power spectrum of mass fluctuations responsible for the lensing.  However, eq. (\ref{VisEq}) above represents an average over the ensemble of all possible lensing fluctuations consistent with a given mass power spectrum, $\Phi_\Sigma$.  In a realistic situation in which the intensity or mutual coherence is averaged over a finite time interval, these quantities will deviate from their ensemble-average values.   The following section considers the magnitude of these deviations and the timescales on which they occur.

\section{Intensity variance and other fourth moments of the wavefield}
In this section we derive expressions for both the intensity variance and the power spectrum of temporal intensity fluctuations.  The latter quantifies the timescale on which fluctuations are expected.  We also consider deviations in the mutual coherence function, $V({\bf r})$, from its mean (ensemble-average) value by calculating both the variance and temporal power spectrum of its fluctuations.  All quantities of interest here are computed by considering the following fourth-order moment of the wave amplitude,
\begin{eqnarray}
\Gamma_4({\bf r}_1,{\bf r}_2) = \langle u({\bf r}) u^*({\bf r} + {\bf r}_2) u({\bf r}+{\bf r}_1) u^*({\bf r} + {\bf r}_1+{\bf r}_2) \rangle.
\end{eqnarray}
This measures the correlation between the wave amplitudes measured by four hypothetical detectors whose locations are depicted in Figure 1.  Various specializations of this quantity yield measures of particular interest to us.  The choice ${\bf r}_1={\bf r}_2=0$ gives $\langle I^2 \rangle$, while allowing ${\bf r}_1 \neq 0$ yields  the correlation in the intensity, $\langle I ({\bf r}_1+{\bf r}') I({\bf r}') \rangle$, between two hypothetical detectors separated by a displacement ${\bf r}_1$.  Temporal intensity variations are calculated by making the replacement ${\bf r}_1 = {\bf v} t$.  The choice ${\bf r}_1 =0, {\bf r}_2\neq 0$ yields the variance in the mutual coherence $V({\bf r}_2)$.  In the most general case, ${\bf r}_1 \neq 0, {\bf r}_2\neq 0$, the quantity $\Gamma_4$ measures the correlation in the mutual coherence between two pairs of receivers separated by a distance ${\bf r}_1={\bf v} t$, with each pair measuring a visibility $V({\bf r}_2)$.    


To compute $\Gamma_4$ we first consider the most general fourth moment of the wave field, which measures the correlation in the fields measured at the four points ${\bf x}_1$, ${\bf x}_2$, ${\bf x}_3$ and ${\bf x}_4$ on the observer's plane,
\begin{eqnarray}
\gamma ({\bf x}_1, {\bf x}_2, {\bf x}_3, {\bf x}_4) &\equiv& \langle \phi({\bf x}_1) \phi^*({\bf x}_2) \phi({\bf x}_3) \phi^*({\bf x}_4)\rangle =
\frac{1}{(2 \pi r_{\rm F}^2)^4} \int d^2{\bf x}_1' d^2{\bf x}_2' d^2{\bf x}_3' d^2{\bf x}_4' 
\exp \left\{\frac{i}{2 r_{\rm F}^2} \left[ ({\bf x}_1' - {\bf x}_1)^2 - ({\bf x}_2' - {\bf x}_2)^2 + ({\bf x}_3' - {\bf x}_3)^2 - ({\bf x}_4' - {\bf x}_4)^2  \right]  \right. \nonumber \\
&\null& \left.  \hskip 5cm + i K \left\langle \int d \bxi \,\Sigma (\bxi) 
\left[ \ln |\bxi -{\bf x}_1' | - \ln |\bxi -{\bf x}_2' | + \ln |\bxi -{\bf x}_3' | - \ln |\bxi -{\bf x}_4' |  \right] \right\rangle\right\}.
\end{eqnarray}
Making the co-ordinate transformation,
\begin{eqnarray}
{\bf R}' &=& {1 \over 4}[{\bf x}_1'+{\bf x}_2' + {\bf x}_3' + {\bf x}_4'] \nonumber \\
{\bf r}_1'&=& {1 \over 2} [{\bf x}_1' + {\bf x}_2' - {\bf x}_3' - {\bf x}_4'] \nonumber \\
{\bf r}_2'&=& {1 \over 2} [{\bf x}_1' - {\bf x}_2' - {\bf x}_3' + {\bf x}_4'] \nonumber \\
{\brho}' &=& {\bf x}_1'-{\bf x}_2' + {\bf x}_3' - {\bf x}_4',
\end{eqnarray}
with a corresponding change of variables for the unprimed coordinates, the fourth moment then becomes,
\begin{eqnarray}
\gamma ({\bf x}_1, {\bf x}_2, {\bf x}_3, {\bf x}_4) &=&
\frac{1}{(2 \pi r_{\rm F}^2)^4} \int d^2{\bf r}_1' d^2{\bf r}_2' d^2{\bf R}' d^2 \brho' \exp \left\{ \frac{i}{r_{\rm F}^2} \left[ ({\bf r}_1'-{\bf r}_1) \cdot ({\bf r}_2-{\bf r}_2') + (\brho - \brho') \cdot ({\bf R} - {\bf R}') \right]  \right\} \nonumber \\
&\null&  \qquad \qquad \qquad \qquad \left\langle \exp \left[ i K \int d \bxi \,\Sigma (\bxi) 
\left( \ln |\bxi -{\bf x}_1' | - \ln |\bxi -{\bf x}_2' | + \ln |\bxi -{\bf x}_3' | - \ln |\bxi -{\bf x}_4' |  \right) \right] \right\rangle. \nonumber \\
\end{eqnarray}
Cast in this form, we apply the simplifying assumption that the gravitational phase, $\psi$, is wide-sense stationary, so that the statistics of the phase differences are independent of the centre co-ordinate, ${\bf R}'$.  The integration over ${\bf R'}$ yields a factor $e^{i {\bf R} \cdot (\brhosm-\brhosm')} \delta (\brho - \brho')$.  Figure 1 shows that the four receivers are arranged in a parallelogram, implying $\brho=0$, and integration over $\brho'$ yields a factor $(2 \pi r_{\rm F}^2)^2$ leaving,
\begin{eqnarray}
\Gamma_4 ({\bf r}_1,{\bf r}_2) &=&
\frac{1}{(2 \pi r_{\rm F}^2)^2} \int d^2{\bf r}_1' d^2{\bf r}_2' \exp \left[ \frac{i}{r_{\rm F}^2} ({\bf r}_1'-{\bf r}_1) \cdot ({\bf r}_2-{\bf r}_2')  \right] \times \left\langle \exp \left[  
i K Z ({\bf r}_1',{\bf r}_2') \right] \right\rangle,  \label{Gamma4}
\end{eqnarray}
where the effect of the gravitational phase delays is embodied in the quantity,
\begin{eqnarray}
Z({\bf r}_1',{\bf r}_2') &=& \int d \bxi \, \ln |\bxi| \, \left[ 
\Sigma \left(\bxi + \frac{{\bf r}_1'+{\bf r}_2'}{2} \right)  -  \Sigma \left(\bxi + \frac{{\bf r}_1'-{\bf r}_2'}{2} \right) +  \Sigma \left(\bxi - \frac{{\bf r}_1'+{\bf r}_2'}{2} \right) -  \Sigma \left(\bxi - \frac{{\bf r}_1'-{\bf r}_2'}{2} \right)\right]. \nonumber \\
\end{eqnarray}
In the following two subsections averages over the phase fluctuations are computed to derive expressions for the variance and power spectrum of both intensity and mutual coherence deviations.

\begin{figure}
\centerline{\psfig{file=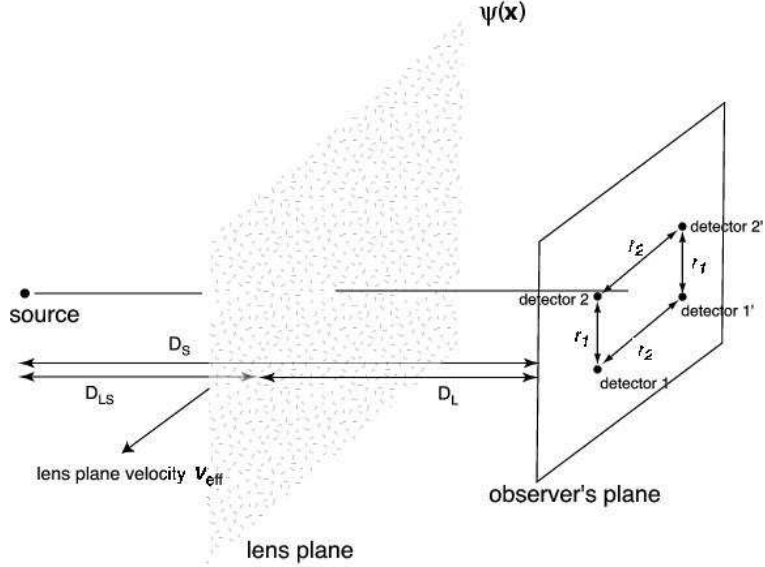,width=10cm}}
\caption{The fourth-order moment of the wave field, $\Gamma_4({\bf r}_1,{\bf r}_2)$ measures the correlation between the two visibilities, each of length ${\bf r}_2$, measured a distance ${\bf r}_1$ apart. Of the four hypothetical dectors depicted, detectors $1$ and $1'$ measure the first visibility, while dectors $2$ and $2'$ measure the second.  The quantity $\Gamma_4(0,{\bf r}_2)$ measures the variance in the visibility $V({\bf r}_2)$.  When the baseline ${\bf r}_2$ is set to zero, so that dector $1$ is co-located with dector $1'$, and dector $2$ is co-located with dector $2'$, the visibility $V({\bf r}_2)$ collapses to just the wave intensity.  The quantity $\Gamma_4({\bf r}_1,0)$ thus measures the autocorrelation of the intensity 
$\langle I({\bf r}+{\bf r}_1) I({\bf r}) \rangle$.  The quantity $\Gamma_4(0,0)$, when all four detectors are at the same location, measures $\langle I^2 \rangle$.}
\end{figure}

\subsection{Small phase perturbations}

When the lensing-induced phase fluctuations are small the fourth moment is well-approximated by  expanding eq. (\ref{Gamma4}) in powers of $K\,Z({\bf r}_1',{\bf r}_2')$,   
\begin{eqnarray}
\Gamma_4({\bf r}_1,{\bf r}_2) &=& \frac{1}{(2 \pi r_{\rm F}^2)^2} \int d^2{\bf r}_1' d^2{\bf r}_2' \exp 
\left\{\frac{i}{r_{\rm F}^2} \left[ ({\bf r}_1'-{\bf r}_1) \cdot ({\bf r}_2-{\bf r}_2')  \right] \right\}  
 \left[ 1 + i K \langle Z({\bf r}_1',{\bf r}_2') \rangle - \frac{K^2}{2} \langle Z({\bf r}_1',{\bf r}_2')^2 \rangle + {\cal O}\left( K^3 \right) \right]. \nonumber \\
\end{eqnarray}
This approximation is valid when the phase delay term $K Z({\bf r}_1',{\bf r}_2')$ is less than unity over the region of integration.  The effective integration area is determined by the region over which the kernel $\exp[({\bf r}_1-{\bf r}_1')\cdot({\bf r}_2-{\bf r}_2')/r_{\rm F}^2]$ varies slowly.  This condition is satisfied when the rms phase delay across a distance of order the Fresnel scale on the lensing plane is less than unity.  We shall see later that for most cases of interest the weak scattering approximation applies under far less restrictive conditions.  

No assumption is made about the statistics of the mass fluctuations except that they are wide-sense stationary and thus that the average mass density is independent of position, so one has $\langle \Sigma({\bf x}) \rangle = \langle \Sigma({\bf x}+{\bf x}') \rangle$ and $\langle Z \rangle =0$.  It is easily shown that the expectation of $Z^2$, written in terms of mass surface density autocorrelation functions, is
\begin{eqnarray}
\langle Z^2({\bf r}_1,{\bf r}_2) \rangle &=& \int d \balpha d \bbeta \, \ln |\balpha| \ln |\bbeta |  \Big{[} 
4 C_\Sigma (\balpha - \bbeta) 
- 2 C_\Sigma (\balpha - \bbeta - {\bf r}_2') - 2 C_\Sigma (\balpha - \bbeta + {\bf r}_2') 
- 2 C_\Sigma (\balpha - \bbeta - {\bf r}_1') - 2 C_\Sigma (\balpha - \bbeta + {\bf r}_1')
	\nonumber \\ &\null& \qquad \qquad 
+ C_\Sigma (\balpha - \bbeta +{\bf r}_1' +  {\bf r}_2') + C_\Sigma (\balpha - \bbeta +{\bf r}_1' -  {\bf r}_2')
+ C_\Sigma (\balpha - \bbeta -{\bf r}_1' +  {\bf r}_2') + C_\Sigma (\balpha - \bbeta -{\bf r}_1' -  {\bf r}_2') 
\Big{]}. \label{Zsqr}
\end{eqnarray}
The relationship between the mass density autocorrelation function and power spectrum in equation (\ref{CtoPhi}) allows $\langle Z^2 \rangle$ to be recast in terms of the power spectrum of mass density fluctuations. 
Subsequent integration over ${\bf r}_1'$ and ${\bf r}_2'$ yields our main expression for the fourth moment in the regime of weak scattering, 
\begin{eqnarray}
\Gamma_4({\bf r}_1,{\bf r}_2) &=& 1 - K^2 \int d^2 {\bf q}  \, q^{-4} \Phi_\Sigma ({\bf q}) \Big{\{} 2 - 2 \cos {\bf q} \cdot {\bf r}_2   -
4 e^{i {\bf q} \cdot {\bf r}_1} \sin^2 \left[ \frac{{\bf q} \cdot ({\bf r}_2 - {\bf q} r_{\rm F}^2 ) }{2} \right]  \Big{\} }.  \label{WeakFour}
\end{eqnarray}
Equivalent results have been derived in other fields in which similar physics applies; it has been investigated extensively by Cronyn (1972) in the field of interplanetary scintillation.  The main difference is that in the present case the phase delays are driven by distortions imposed by matter on the geometry of spacetime, whereas for interplanetary scintillation phase delays are related to density fluctuations by the plasma dispersion relation.  

Equation (\ref{WeakFour}) is the starting point for computing all quantities of interest in the weak scattering limit.  In this section expressions for all these quantities are listed, but their explicit evaluation for given mass distributions is deferred to the next section.  The most obvious quantity is the total variance of the intensity fluctuations.  Since $\Gamma_4(0,0) = \langle I^2 \rangle$ the intensity variance is
\begin{eqnarray}
 \langle \Delta I^2 \rangle &=& \Gamma_4(0,0)-1= 4 K^2 \int d^2{\bf q} \, q^{-4} \Phi_\Sigma({\bf q}) \sin^2 \left( \frac{q^2 r_{\rm F}^2}{2}\right). \label{DeltaIsqr} \label{Ivar}
\end{eqnarray}
The timescale on which the intensity fluctuations occur is described by the power spectrum of the intensity fluctuations.  It was mentioned above that the fourth moment gives the autocorrelation of the intensity fluctuations between points separated by a vector ${\bf r}_1$.  The Fourier transform of this moment is used to obtain the power spectrum of intensity fluctuations across the observer's plane:  
\begin{eqnarray}
W_I({\bf q}) &=&  \int d^2{\bf r}_1 \exp[-i {\bf q} \cdot {\bf r}_1] \, \Gamma_4({\bf r}_1,0).
\end{eqnarray}
The integral of $W_I({\bf q})$ over all ${\bf q}$ of course yields the variance in the intensity.
In the weak scattering limit the spatial power spectrum of intensity fluctuations is derived from the fourth order moment by noting that eq. (\ref{WeakFour}) is already written in terms of a Fourier transform.  Setting ${\bf r}_2=0$ and Fourier transforming $\Gamma_4({\bf r}_1,0)$ with respect to ${\bf r}_1$ we see that the power spectrum of intensity fluctuations is, 
\begin{eqnarray}
W_I({\bf q}) =  4 K^2 q^{-4} \Phi_\Sigma ({\bf q}) \sin^2 \left[ \frac{ q^2 r_{\rm F}^2  }{2} \right]. \label{WI}
\end{eqnarray}
Wave effects become particularly important for wavenumbers $q$ greater than $r_{\rm F}^{-1}$, where the power spectrum oscillates with a frequency determined by the Fresnel scale.  On the other hand, the power spectrum is an unweighted measure of the power spectrum of mass surface density fluctuations for wavenumbers $q$ less than $r_{\rm F}^{-1}$.  In particular, one has $W_I({\bf q}) \approx  K^2 r_{\rm F}^4 \Phi_\Sigma ({\bf q})$ for  $q \lesssim r_{\rm F}^{-1}$.  In this regime the power spectrum is also independent of observing frequency, with the frequency dependence of the Fresnel scale being offset by the dependence of $K$.  At higher observing frequencies wave effects become less important, the Fresnel scale becomes smaller, and the regime of applicability of the approximation $q < r_{\rm F}^{-1}$ means that wave effects are important only for progressively smaller length scales.

In practice it is difficult to measure the power spectrum of intensity fluctuations across a two-dimensional observing plane.  A more practical observable quantity is the power spectrum of intensity variations experienced by a single detector.  In general the lensing material (e.g. stars, gas and dark matter) will be in motion relative to the line of sight to the source.  If the motions of the individual lensing constituents are assumed to be negligible relative to the bulk velocity of the lensing plane across the line of sight, the so-called frozen-screen approximation, the effective lensing velocity is,
\begin{eqnarray}
{\bf v} = {\bf v}_{\rm screen} - \left[ {\bf v}_{\rm Earth} \left(1 -\frac{D_L}{D_S} \right) + {\bf v}_{\rm src} \left( \frac{D_S}{D_L} \right) \right],
\end{eqnarray}
where ${\bf v}_{\rm Earth}$, ${\bf v}_{\rm screen}$ and ${\bf v}_{\rm src}$ are, respectively, the velocities of the Earth, lensing plane and the source transverse to the line of sight.  The power spectrum of temporal intensity fluctuations is obtained from the fourth-order moment by equating spatial and temporal displacements according to the rule ${\bf r}  \rightarrow {\bf v} t$ to yield\footnote{Since  coordinates on the observing plane are scaled (see the text below eq. \ref{uGrav}), $\omega$ is also scaled.  The replacement $\omega \rightarrow (D_{LS}/D_S)\omega$ converts to observational quantities.  This replacement is insubstantial in most cases under consideration since the modifying factor is of order unity.},
\begin{eqnarray}
W_I (\omega) = \frac{2 \pi}{v} \int dt \, e^{i \omega t} \, \Gamma_4({\bf v} t, 0). \label{Womegadefn}
\end{eqnarray}
It is often the case that the power spectrum of mass fluctuations is an isotropic function of ${\bf q}$, so that the fourth moment is also an isotropic function of ${\bf r}_1$.  Without loss of generality the velocity can then be oriented along the $x$--axis ${\bf v}=(v,0)$, and the power spectrum of temporal intensity fluctuations is thus,
\begin{eqnarray}
W_I (\omega) = \frac{8 \pi K^2 }{v} \int dq \left[ \frac{\omega^2}{v^2} + q^2 \right]^{-2} \Phi_\Sigma \left(\frac{\omega}{v},q \right) \sin^2 \left[ \frac{\left( \frac{\omega^2}{v^2}+ q^2 \right) r_{\rm F}^2}{2} \right]. \label{Womega}
\end{eqnarray}

Both the variance and power spectrum of the mutual coherent fluctuations are derived in a similar manner to the intensity fluctuations.   In subtracting out the mean mutual coherence to determine the variance, it is useful to recognise that the first three terms in eq. (\ref{WeakFour}) just represent the square mean mutual coherence $V({\bf r}_2)$ in the small perturbation limit:
\begin{eqnarray}
\langle V({\bf r}_2)\rangle^2 = \exp[-D_\phi({\bf r}_2)/2]^2 = 1 - 2 K^2 \int d^2{\bf q}\, q^{-4} \Phi_\Sigma({\bf q})   [1- \cos( {\bf r}_2 \cdot {\bf q}) ] + {\cal O}(K^4).  
\end{eqnarray}
The exponential function is expanded in the same way that it was linearised when making the weak scattering approximation for the fourth order moment.
Making the separation $V^2 = (\bar{V} + \delta V)^2 = \delta V^2 + \bar{V}^2$, the fluctuation in the mutual coherence on a baseline ${\bf r}_2$ is $\Delta V^2(0,{\bf r}_2) = \Gamma(0, {\bf r}_2) - \langle V\rangle^2$,
\begin{eqnarray}
\langle \Delta V^2({\bf r}_2) \rangle = 4K^2 \int d^2{\bf q} \, q^{-4} \, \Phi_\Sigma ({\bf q}) 
\sin^2 \left[ \frac{{\bf q} \cdot ({\bf r}_2 - {\bf q} r_{\rm F}^2) }{2}\right]. \label{VisVar}
\end{eqnarray}
The power spectrum of mutual coherence fluctuations measured on a baseline ${\bf r}_2$ is 
\begin{eqnarray}
W_{\Delta V} (\omega; {\bf r}_2) &=&  \frac{8 \pi K^2}{v} \int dq \left[ \frac{\omega^2}{v^2} + q^2 \right]^{-2} 
\Phi_\Sigma \left( \frac{\omega}{v},q \right) 
\sin^2 \left[ \frac{\frac{\omega}{v} r_{2x}  + q r_{2y} - \left(  \frac{\omega^2}{v^2}+ q^2 \right) r_{\rm F}^2}{2} \right],
\label{VisPow}
\end{eqnarray}
where the mass fluctuation spectrum is again assumed to be an isotropic function of ${\bf q}$.

\subsection{Large phase perturbations}
The regime of strong scattering applies when $K Z({\bf r}_1',{\bf r}_2') \gtrsim 1$ over the effective region of integration in ${\bf r}_1'$ and ${\bf r}_2'$.  In order to compute averages over the phase in this regime, slightly more restrictive assumptions must be made about the statistical properties of the mass density fluctuations than is necessary in the regime of weak scattering.  

The gravitational phase delay is assumed here to be normally distributed.  
This is the case if the underlying mass surface density fluctuations are themselves normally distributed.  Averages over the phase fluctuations are performed using the fact that, for a Gaussian random variable, $\delta x$, with zero mean, the average of $\langle \exp[i\,\delta x] \rangle$ is $\exp [-\langle \delta x^2 \rangle /2]$.  This implies that the average over phase fluctuations, $\langle \exp[i K Z] \rangle = \exp[-K^2 \langle Z \rangle^2/2]$, and the fourth moment of the wave amplitude is
\begin{eqnarray}
\Gamma_4({\bf r}_1,{\bf r}_2) &=& \frac{1}{(2 \pi r_{\rm F}^2)^2} \int d^2{\bf r}_1' d^2{\bf r}_2' \exp 
\left[ \frac{i ({\bf r}_1'-{\bf r}_1) \cdot ({\bf r}_2-{\bf r}_2')}{r_{\rm F}^2} 
- D_\psi({\bf r}_1') - D_\psi({\bf r}_2') + \frac{D_\psi({\bf r}_1'+{\bf r}_2')}{2} + \frac{ D_\psi({\bf r}_1'-{\bf r}_2')}{2}   \right] . \nonumber \\ \label{Gamma4Strong} 
\end{eqnarray}
Unfortunately, the form of eq. (\ref{Gamma4Strong}) in the strong scattering limit is not amenable to further simplification without explicitly specifying the form of the phase structure function.  We thus note that the intensity variance is $\langle \Delta I^2 \rangle = \Gamma_4(0,0) -1$ while the variance in the mutual coherence on a baseline ${\bf r}_2$ is $\langle \Delta V^2 ({\bf r}_2) \rangle = \Gamma_4(0,{\bf r}_2) - \exp[-D_\psi ({\bf r}_2)]$.

The spatial power spectrum of the mutual coherence fluctuations across the observer's plane is obtained by making the replacement ${\bf q} =({\bf r}_2' - {\bf r}_2)/ r_{\rm F}^2$ in eq. (\ref{Gamma4Strong}), which allows the power spectrum to be identified as,
\begin{eqnarray}
W_{\Delta V}({\bf q};{\bf r}_2) = \int d^2{\bf r}_1' \exp \left[ -i {\bf q} \cdot {\bf r}_1' - D_\psi({\bf r}_1') 
- D_\psi({\bf r}_2+{\bf q} r_{\rm F}^2) + \frac{D_\psi({\bf r}_1'+{\bf r}_2+{\bf q} r_{\rm F}^2)}{2} + \frac{ D_\psi({\bf r}_1'-{\bf r}_2- {\bf q} r_{\rm F}^2)}{2}   \right]. \label{WDeltaVStrongSpatial}
\end{eqnarray}
The corresponding power spectrum of the temporal intensity fluctuations is obtained by setting 
${\bf r}_2=0$ in eq. (\ref{WDeltaVStrongSpatial}):
\begin{eqnarray}
W_{I} (\omega) = \frac{1}{2 \pi \, v} \int d^2{\bf r}_1' dq 
\exp \left[ -i {\bf q} \cdot {\bf r}_1' - D_\psi({\bf r}_1') 
- D_\psi({\bf q} r_{\rm F}^2) + \frac{D_\psi({\bf r}_1'+{\bf q} r_{\rm F}^2)}{2} + \frac{ D_\psi({\bf r}_1'- {\bf q} r_{\rm F}^2)}{2}   \right], \label{WDeltaVStrongTemporal}
\end{eqnarray}
where the mass density power spectrum is assumed to be an isotropic function of ${\bf q}=(\omega/v,q)$.
Expressions for the intensity and mutual coherence fluctuations in the strong perturbation regime are only amenable to further simplification once the phase fluctuations have been specified; these are discussed in Appendix \ref{AppCstrong}

Throughout the remainder of this paper we shall be mostly concerned with lensing in the limit of weak phase perturbations.  It is shown in Sect.\,\ref{Discussion} that the large phase perturbation regime requires such large mass fluctuations as to be unlikely to apply even under the most extreme circumstances.


\section{Lensing by a power law mass spectrum} 


In this section we examine the intensity fluctuations caused by a power spectrum of mass surface density fluctuations.  This plausibly applies to lensing by small scale structure in cold dark matter if is clustered heirarchically on small scales.  The power spectrum of dark matter fluctuations on very small scales is unknown, but it is not unreasonable to expect that it may follow a power law to sub-pc scales (e.g. Diemand, Moore \& Stadel 2005). (The alternate possibility is that the matter is distributed in discrete clumps; lensing by such a distribution is examined in Sects.\,5--6.)  In the CDM paradigm the power spectrum of mass fluctuations follows a $P(k) \propto k^{4-2\alpha}$ spectrum on the small spatial scales of interest here (Ch.\,4 Padmanabhan 1993), where $\alpha=3/2$ corresponds to an initially scale-invariant spectrum.   The spectrum of fluctuations cuts off at the free-streaming scale (Padmanabhan 1993; Peacock 1999) which, for dark matter particles of rest mass energy $E_m$, occurs at a spatial scale
\begin{eqnarray}
\lambda_{\rm fs} = 0.005 \left( \frac{E_m}{1\,{\rm GeV}c^{-2}} \right)^{-4/3} \,{\rm pc}.  
\end{eqnarray}
For cold ($E_m> 1\,$GeV$c^{-2}$) dark matter particles the is well below the scales typically probed by gravitational scattering for wave frequencies $\nu \ga 1$\,Hz.  

Notwithstanding the distribution of dark matter on small scales, there are two additional reasons to consider lensing by a power law spectrum of mass fluctuations.  The first is that it applies to lensing by any astrophysical turbulent plasma or gas.  It is now well established that both the ionized interstellar medium of our Galaxy and the HI in our and nearby galaxies follow power law spectra over a large range of spatial scales which encompasses those of interest here, and there is every reason to expect this to apply to other systems (e.g. Armstrong, Rickett \& Spangler 1995; Dickey et al. 2001; Stanimirovic \& Lazarian 2001; Braun 1999).  Although the lensing contribution from gas is likely to be small under most circumstances, it is nonetheless interesting to consider its contribution from a formal standpoint.  It may also apply to lensing by a turbulent proto-galaxy. The second reason is that an examination of the effects of lensing from a power law mass spectrum provides a useful test-bed in which to develop an intuitive understanding of lensing in the long-wavelength regime applicable to gravitational radiation which can then be applied to more complicated lensing situations.  We return to this point at the end of this section.

We consider mass fluctuations with the following power law distribution,  
\begin{eqnarray}
\Phi_\Sigma ({\bf q}) = Q_0 \left\{
 \begin{array}{ll}
q_{\rm min}^{-\beta}, & q_0 < q < q_{\rm min}, \\
q^{-\beta} & q_{\rm min}, < q < q_{\rm max}, \\
0, & q > q_{\rm max}. \\
\end{array} 
\right. \label{MassPwr}
\end{eqnarray}
This simple description is sufficient for all cases of interest here.  Over the range of scales pertinent to lensing in the present case the mass fluctuations are expected to be described by a single power law index.  One can, of course, prescribe a more complicated spectrum (e.g. by changing the behaviour at the inner and outer scales), but only at the expense of extra complexity.  We see from the results below that such extra complexity is unwarranted in most cases.  The spectrum flattens to a constant at an inner wavenumber $q_{\rm min} \equiv L_0^{-1}$, where $L_0$ is interpreted as the outer scale of the density fluctuations.   For shallower power spectra, $\beta<2$, the outer scale of the power spectrum, $q_{\rm max} \equiv l_0^{-1}$, must also be specified to ensure that the mass density variance is finite.  A cutoff scale $q_0$ is also introduced to ensure the convergence of certain integrals evaluated below, but it will become apparent that this scale is physically unimportant\footnote{Note that the form of the power spectrum differs slightly to that considered in M04, where the power spectrum of mass density (not mass surface density), $\Phi_\rho$, fluctuations was specified directly. The previous formulation is too simple for treating intensity fluctuations, as it included no power beyond the outer scale.}.  Expressed in terms of the mass surface density variance, the constant of proportionality $Q_0$ is,
\begin{eqnarray}
Q_0 = \langle \Delta \Sigma^2 \rangle \left\{ \begin{array}{ll}
\frac{2 -\beta}{2 \pi}  q_{\rm max}^{\beta-2}, & \beta < 2, \\
\frac{\beta-2}{\beta \pi} q_{\rm min}^{\beta-2} ,  & \beta >2,
\end{array}  \right. 
\end{eqnarray}
where $q_{\rm min} \ll q_{\rm max}$ is assumed. 
For $q_0 \ll q_{\rm min}$ the structure function takes the following form (see Appendix \ref{DsAppendix}),
\begin{eqnarray}
D_\psi ({\bf r}) = 4 \pi Q_0 K^2 \left\{ r^2 \frac{q_{\rm min}^{-\beta}}{4} \left[\log \left( \frac{q_{\rm min}}{q_0} \right) - \frac{1}{\beta} \right]  
- r^{2+\beta} \frac{\Gamma \left(-1 - \frac{\beta}{2} \right)}{2^{3+\beta} \Gamma \left( 2+ \frac{\beta}{2} \right)} 
- r^4 \frac{\beta q_{\rm min}^{2-\beta}}{128 (\beta-2)}  + {\cal O}\left( r^6 q_{\rm min}^{4-\beta} \right) 
\right\},  \qquad 0 < \beta < 4 . \nonumber \\ \label{Ds}
\end{eqnarray}
It is convenient to cast this in the form,
\begin{eqnarray}
D({\bf r})  = \left(\frac{r}{r_{\rm diff}} \right)^2 -  \left( \frac{r}{r_0} \right)^{{\rm min}[2+\beta,4]}, \qquad 0 < \beta < 4, \quad r < q_{\rm min}^{-1},
\end{eqnarray}
where only the two most important terms are retained.  This separation is useful because inspection of $Z^2({\bf r}_1,{\bf r}_2)$ (cf. eq. (\ref{Zsqr})) shows that the $r^2$ terms make no contribution to the intensity fluctuations.  The term proportional to $r^2$ dominates for $r \ll q_{\rm min}^{-1}$, so the transverse separation between points on the lensing plane over which the rms phase changes by one radian is
\begin{eqnarray}
r_{\rm diff} = \frac{1}{K \langle \Delta \Sigma^2 \rangle^{1/2} }\sqrt {\frac{2}{\pi \left[ \log \left( \frac{q_{\rm min}}{q_0}\right) - \frac{1}{\beta} \right]} } \left\{ \begin{array}{ll}
q_{\rm max} \left( \frac{q_{\rm max}}{q_{\rm min}} \right)^{\beta/2}  \frac{1}{\sqrt{\beta-2}}, & 0 < \beta < 2 \\
q_{\rm min} \frac{1}{\sqrt{2-\beta}}, & \beta > 2 . \\
\end{array} \right.
\end{eqnarray}
However, since the $r^2$ term makes no contribution to intensity fluctuations, the salient quantity when considering intensity fluctuations is $r_0$, which defines the length scale over which phase fluctuations that contribute to $Z^2$ change by one radian rms,
\begin{eqnarray}
r_0 = \left\{ \begin{array}{ll}
\left[\frac{q_{\rm max}^{2-\beta} 2^{2+\beta} \Gamma \left(2+\frac{\beta}{2} \right)}{ (2-\beta) 
\langle \Delta\Sigma^2 \rangle K^2 \Gamma \left(-1-\frac{\beta}{2} \right) } \right]^{1/(\beta+2)}, 
	&  0<\beta < 2 \\
 2^{5/4} |K|^{-1/2} \langle \Delta \Sigma^2\rangle^{-1/4}, &  2 <\beta<4 .
\end{array}
\right. \label{s0defn}
\end{eqnarray}
It is the magnitude of this quantity, and not $r_{\rm diff}$, which discriminates between weak and strong scintillation regimes.  Strong scintillation corresponds to the regime $r_0 \ll r_{\rm F}$.  Since it is far harder to satisfy the criterion $ r_0 \ll r_{\rm F}$ than $r_{\rm diff} \ll r_{\rm F}$, the scattering can be often be classified as weak even when the relative phase delay between two points separated by a Fresnel scale on the lensing plane greatly exceeds unity.  Thus the weak scattering condition holds over a larger range of scattering conditions than in other applications of scattering physics (see, e.g., Narayan 1992).

\subsection{Intensity fluctuations}
The magnitude of intensity fluctuations in the regime of weak scattering is evaluated by inserting eq. (\ref{MassPwr}) in eq. (\ref{DeltaIsqr}), 
\begin{eqnarray}
\langle \Delta I^2 \rangle = 4 K^2 Q_0 \left[ 2 \pi q_{\rm min}^{-\beta} \int_0^{q_{\rm min}} dq\, q^{-3}  \sin^2 \left( \frac{q^2 r_{\rm F}^2}{2} \right) 
+ 
2 \pi \int_{q_{\rm min}}^{\infty} dq\, q^{-3-\beta} \sin^2 \left( \frac{q^2 r_{\rm F}^2}{2} \right) 
\right].
\end{eqnarray}
In the limit $q_{\rm min} r_{\rm F} < 1$ the modulation amplitude depends on the steepness of the mass power spectrum, with a change in the nature of the solution at $\beta=2$.
By recasting $K^2 Q_0$ in terms of $r_0$, the scale length over which phase fluctuations responsible for focusing vary, we see that the magnitude of the intensity fluctuations in the regime of weak scattering depends only on the ratio $r_{\rm F}/r_0$,
\begin{eqnarray}
\langle \Delta I^2 \rangle = \left\{
\begin{array}{ll}
2^{2+\beta} \Gamma \left( 2+ \frac{\beta}{2} \right) \sin \left( \frac{\pi \beta}{4} \right) \left( \frac{r_{\rm F}}{r_0} \right)^{2+\beta}, & 0 < \beta < 2, \\
2^5 \left( \frac{r_{\rm F}}{r_0}\right)^{4}, & 2 < \beta < 4. \\
\end{array}
\right. \label{WeakScatRes}
\end{eqnarray}
Furthermore, the condition for weak scattering, $r_{\rm F} < r_0$, implies that the intensity fluctuations in this regime are less than or comparable to unity.  The amplitude of the intensity variations may be understood physically by regarding the phase delay imparted by the mass distribution as a perturbation to the phase curvature over the coherence region, of radius $r_{\rm F}$, that contributes to the intensity on the observer's plane.   The rms intensity deviation is $\approx D_{\rm eff}/f$, where $D_{\rm eff} = D_{L} D_{LS}/D_S$ is regarded as the effective distance of the observer behind the lensing plane and $f$ is the focal length of the phase perturbation.  The focal length of a lens that retards the wavefront by a phase $\Delta \phi$  over a distance $r$ from the lens centre is $r^2 k/\Delta \phi$.   The rms phase curvature across the Fresnel scale is $D_\psi (r_{\rm F})^{1/2}$, where we consider only the non-quadratic contribution to the phase structure function because the terms $r^2$ do not give rise to intensity fluctuations.  The rms intensity deviation is thus $(r_{\rm F}/r_0)^{{\rm min}[1+\beta/2,2]}$.

For very low frequency waves, $\nu \sim 10^{-8}\,$Hz, it is conceivable that the Fresnel scale becomes comparable to or even exceeds the outer scale of the power spectrum.  For completeness we include the expression for the intensity variance in the limit $q_{\rm min} r_{\rm F} > 1$,
\begin{eqnarray}
\langle \Delta I^2 \rangle = \pi^2 K^2 Q_0 q_{\rm min}^{-\beta} r_{\rm F}^2 + {\cal O}(q_{\rm min}^{-2-\beta}).
\end{eqnarray}
The intensity variance in this instance depends even more strongly on the outer scale of the power spectrum than in the $q_{\rm min} r_{\rm F} < 1$ case.

The foregoing discussion is related to the amplitude of the intensity fluctuations.  However, several other observables are of interest, in particular the power spectrum of intensity fluctuations and the amplitude and power spectrum of fluctuations in the mutual coherence.  These quantities are considered in detail in Appendix \ref{AppC}.  

Although the above results are derived using a specific model of the mass fluctuations, they lead to a physical understanding of the lensing which should apply more generally.  The amplitude of the intensity fluctuations has a simple physical interpretation.  It is governed by the phase curvature induced by the mass fluctuations across the coherence scale, which is set by the Fresnel length, and intensity fluctuations arise due to the focussing and defocussing of radiation across this scale.  Accordingly, most of the fluctuation power occurs on timescales $\ga r_{\rm F}/v$, the time on which the line of sight traverses the Fresnel scale (see Appendix \ref{AppC} for details).  The utility of these results is demonstrated in the context of stellar microlensing below.  

\section{The contribution from stars and other discrete objects}

\subsection{Single lensing population}

Equations (\ref{Ivar}) and (\ref{Womega}) demonstrate that the effect of lensing can be ascertained directly from the power spectrum of the mass fluctuations. Here we consider the effect of stellar microlensing by writing the power spectrum of the mass surface density in terms of the stellar mass profile and the spatial distribution of stars across the lensing plane.  A simple prescription for determining the power spectrum for a collection of stars of identical masses was introduced in Macquart (2004), founded upon the work of Melrose (1996), but here we generalize the formalism to take into account the clustering properties of the stars and a distribution of stellar masses.  
Consider first the case of lensing by a collection of $N$ identical stars with positions ${\bf r}_i$ on the lensing plane.  The mass surface density takes the form,
\begin{eqnarray}
\Sigma({\bf r}) = \sum_{i=0}^{N_M} f_M({\bf r} - {\bf r}_i), \label{MassForm}
\end{eqnarray}
where $f_M({\bf r}) $ is the projected density mass of a star of total mass $M$ centred on the origin.
One is free to substitute any particular form for the stellar mass density profile, but for our purposes it suffices to approximate the stars as constant-density objects of radius $R$.  The projected surface density associated with such objects is $f_M({\bf r}) = (3M/2 \pi R^3) \sqrt{R^2 - r^2}$, and the power spectrum of the mass surface density is
\begin{eqnarray}
\tilde f_M({\bf q}) = \frac{3 M}{R^3} \left[ \frac{\sin (q R) - q R \cos (q R)}{q^3} \right] 
=  M \left( 1 - \frac{R^2 q^2}{10}  \right) + {\cal O}(q^4 R^4).  \label{fM}
\end{eqnarray}
The power spectrum of the surface density is constant for $q< \sqrt{5} R^{-1}$.

The power spectrum of mass density fluctuations on the lensing plane which is related to the covariance of the mass surface density distribution,
\begin{eqnarray}
C_\Sigma ({\bf r}) = \frac{1}{(2 \pi)^2} \int d^2{\bf q}\, e^{i {\bf q} \cdot {\bf r}} \Phi_\Sigma ({\bf q}).
\end{eqnarray}
The covariance, $C_\Sigma({\bf r})$, of the mass distribution is computed by determining the Fourier transform of the mass surface density profile as follows:
\begin{eqnarray}
\langle \Sigma ({\bf r}' + {\bf r}) \Sigma ({\bf r}') \rangle &=&  
\frac{1}{\cal A} \int d^2{\bf r}' \, \Sigma({\bf r}' + {\bf r}) \Sigma ({\bf r}') 
= \frac{1}{(2 \pi)^2 {\cal A}} \int d^2{\bf q} \, e^{i {\bf q} \cdot {\bf r} } | \tilde \Sigma ({\bf q}) |^2,
\label{StarCorrelation}
\end{eqnarray}
where ${\cal A}$ is the area of the lensing plane.  Since only fluctuations in the mass surface density across the lensing plane give rise to intensity fluctuations, one should strictly subtract the contribution of the mean density, $\langle \Sigma \rangle$, from the power spectrum.   However, since the mean is obviously independent of position across the lensing plane it contributes only at the single point ${\bf q}=0$.  Thus the subtraction of the mean density does not affect our results and the correction is subsequently ignored.

The average power spectrum of the mass surface density is 
computed by separating the $N$ self ($i=j$) contributions from the $N(N-1)$ cross-term ($i \neq j$) contributions and identifying $\sigma = N/{\cal A}$ as the mean stellar surface density:
\begin{eqnarray}
 \Phi_\Sigma ({\bf q})   = \sigma  | \tilde f_M({\bf q}) |^2 +
\sigma (N-1) | \tilde f_M({\bf q}) |^2 \left\langle e^{i {\bf q} \cdot ({\bf r}_i - {\bf r}_j)} \right\rangle . \label{PwrSum}
\end{eqnarray}
The average over positions ${\bf r}_i$ and ${\bf r}_j$ appearing in the last term vanishes if the masses are distributed randomly over all space.  However, this average is in general non-zero if the stars are clustered.   Now if the distribution of stars is statistically homogeneous (i.e. it is wide-sense stationary), the joint distribution of object positions, $p({\bf r}_i,{\bf r}_j)$, depends only on the difference, $\Delta{\bf r} = {\bf r}_i-{\bf r}_j$, and is independent of the average position ${\bf s} = ({\bf r}_i+{\bf r}_j)/2$. Writing $p({\bf r}_i, {\bf r}_j)$ as a function of $\Delta {\bf r}$ only, the average over object positions is, 
\begin{eqnarray}
\left\langle e^{i {\bf q} \cdot ({\bf r}_i - {\bf r}_j)} \right\rangle &=& \frac{1}{{\cal A}^2} \int d^2\Delta {\bf r}
 \, d^2 \Delta {\bf s} \, p(\Delta {\bf r}) e^{i {\bf q} \cdot \Delta {\bf r}} 
= \frac{\tilde p_2({\bf q})}{\cal A} . \label{PosnAvg}
\end{eqnarray}
The function $p({\bf r})$ is regarded as the {\it fraction} of objects whose separation is between ${\bf r}$ and ${\bf r}+d{\bf r}$.
Placing these averages back into equation (\ref{PosnAvg}),  
the power spectrum of surface density fluctuations takes the simple form,
\begin{eqnarray}
\Phi_\Sigma ({\bf q}) &=&  \sigma  | \tilde f_M({\bf q}) |^2 \left[  1 + \sigma \tilde p_2({\bf q}) \right] 
\label{SingleStarPhi} 
\end{eqnarray}
This spectrum can be inserted directly into equations (\ref{Ivar}) \& (\ref{Womega}) to determine the characteristics of intensity fluctuations due to lensing by an ensemble of identical stars.  A specific model for stellar clustering is introduced below in which eq. (\ref{SingleStarPhi}) is evaluated explicitly.

\subsection{Lensing by stars of different masses}

The generalization to stars of multiple masses is made by dividing up the stars into $N_{\rm pop}$ mass bins with the $N_{M_j}$ stars within each mass bin, $j$, located at positions ${\bf r}_{i,j}$.  The corresponding mass surface density takes the form 
\begin{eqnarray}
\Sigma ({\bf r}) = \sum_j^{N_{\rm pop}} \sum_i^{N_{M_j}} f_{M_j} \left( {\bf r}-{\bf r}_{i,j} \right).
\end{eqnarray}
We proceed as above in computing the power spectrum of mass density fluctuations, which is now divided into three distinct sets of terms, (i) the self terms within each mass bin, (ii) cross terms within each mass bin (i.e. the contribution from stars of identical mass but different positions) and, (iii) cross terms from stars of different mass.  Terms of the kind (i) and (ii) appeared in the single-mass case above, but the contribution from (iii) is new. The mass power spectrum is 
\begin{eqnarray}
 \Phi_\Sigma ({\bf q})  &=& \frac{1}{\cal A}
\sum_k^{N_{\rm tot}} \left\vert \tilde f_{M_k}({\bf q}) \right\vert^2  \left[ N_{M_k}  + N_{M_k} (N_{M_k} -1)\left\langle e^{i {\bf q} \cdot ({\bf r}_{i,k} - {\bf r}_{j,k}) } \right\rangle  \right]
+ 
\frac{1}{\cal A} \sum_{k \neq l}^{N_{\rm tot}} N_{M_k} N_{M_l} \tilde f_{M_k}({\bf q}) \tilde f_{M_l}^* ({\bf q}) \left\langle e^{i {\bf q} \cdot ({\bf r}_{i,k} - {\bf r}_{j,l}) }\right\rangle. \nonumber \\ \label{AllTerms}
\end{eqnarray}
In performing the average over star positions drawn from different mass bins we introduce the function $p_{kl} ({\bf r})$, which measures the fraction of stars, of mass $M_k$ and $M_l$, separated by a displacement ${\bf r}$.  Thus the mass power spectrum takes the final form
\begin{eqnarray}
\Phi_\Sigma ({\bf q}) = \sum_k^{N_{\rm tot}} \sigma_{M_k} \left\vert \tilde f_{M_k}({\bf q}) \right\vert^2 \left[1 + \sigma_{M_k} \tilde p_{kk} ({\bf q}) \right]  
+ \sum_{k \neq l}^{N_{\rm tot}} \sigma_{M_k} \sigma_{M_l} \tilde f_{M_k}({\bf q}) \tilde f_{M_l}^*({\bf q}) \tilde p_{kl}({\bf q}). 		\label{PhiCross}
\end{eqnarray}
Two properties of the model must be specified in order to compute the mass density power spectrum, (i) the mass distributions of the individual stars and (ii) the clustering properties.  The effect of the latter is embodied in the probability distribution of interstellar separations, $p_{kl}({\bf r})$.  Obviously if clustering is unimportant (i.e. the stars are distributed uniformly) this probability distribution is independent of ${\bf r}$ and all the cross-terms in eq. (\ref{PhiCross}) are then proportional [i.e. $\tilde p_{kl} ({\bf q}) \propto \delta ({\bf q})$] and are thus unimportant.  The following subsection introduces a simple model to evaluate the effect of clustering on the mass power spectrum.  The mass distribution is addressed subsequently.

\subsection{Stellar clustering}
A statistical description of stellar clustering properties is important when the stellar surface density is high.  Terms related to stellar clustering scale quadratically with the stellar surface density in the mass power spectrum, whereas uniformly distributed stars make a contribution that is only linearly proportional to the stellar density.  Clustering properties enter into the power spectrum through the contribution of the cross terms $\sigma_{M_k}(N_{M_k}-1) \langle \exp[i {\bf q} \cdot ({\bf r}_{i,k} - {\bf r}_{j,k}) ]\rangle$ and $\sigma_{M_k}(N_{M_l}-1) \langle \exp[i {\bf q} \cdot ({\bf r}_{i,k} - {\bf r}_{j,l}) ]\rangle$ in eq. (\ref{AllTerms}) above.  The first cross term is due to the clustering between stars of identical mass, while the second represents clustering between stars of different mass.  We make the simplifying assumption that stellar clustering properties depend only on stellar mass, so that stars of identical mass all possess identical clustering characteristics.

The effect of clustering between stars of identical type is determined from the probability distribution of stellar separations between all pairs of stars on the lensing plane.  Clustering information is embedded in the two-point correlation function, $\xi ({\bf r})$, in which the differential probability of finding two stars within areas $dA_1$ and $dA_2$ respectively is 
\begin{eqnarray}
dP = \sigma^2  [1+ \xi({\bf x})] dA_1 dA_2,
\end{eqnarray}
where $\sigma$ is the stellar surface density.  It is unclear exactly how to parameterize stellar clustering, so we explore the following simple form which is found to apply to clustering between galaxies in cosmological surveys (Peebles 1980a):
\begin{eqnarray}
\xi({\bf r}) = \left( \frac{r}{r_0} \right)^{-\gamma}, \label{xi}
\end{eqnarray}
where $r_0$ is regarded as the clustering amplitude to be determined either empirically or using simple physical arguments.  Despite possessing the dimensions of length, it is not useful to interpret $r_0$ as a typical clustering length because the power-law form assumed for $\xi$ for above is scale-free. 
Given the location of a star, the number of stars surrounding it within a radius $R$ is, 
\begin{eqnarray}
N(r<R) = \sigma \int_{\cal A} d^2{\bf r} \, [1+\xi({\bf r}) ] = \pi R^2 \sigma 
\left[1 + \frac{2}{2-\gamma} \left( \frac{R}{r_0} \right)^{-\gamma} \right] .
\end{eqnarray}
For $0<\gamma < 2$ the form of $\xi({\bf r})$ in eq. (\ref{xi}) suffices to provide a physically acceptable description of stellar clustering over all possible separations, because the function $N(R)$ is well-behaved over the entire range $0<R< \infty$.  It satisfies the requirement that the excess probability of finding a star goes to zero at large separations provided $\gamma>0$.  More importantly, it also satisfies the criterion that $N(R)$ is finite at small $R$ provided $\gamma < 2$, which removes the necessity of  imposing an arbitrary inner cutoff to limit the value of $\xi({\bf r})$ at small $r$.

The number of stars within the thin annulus bounded by ${\bf r} + \Delta {\bf r}$ and ${\bf r}$ is, 
\begin{eqnarray} 
\frac{\partial N}{\partial r} \Delta r = 2 \pi r \sigma \left[ 1+ \left( \frac{r}{r_0} \right)^{-\gamma} \right]\,\Delta r.
\end{eqnarray}
This quantity is directly proportional to the probability distribution of object separations.  
The probability distribution of object separations across the lensing plane is normalized by requiring that the total probability of finding the second object somewhere within ${\cal A}$ is unity, yielding,
\begin{eqnarray}
p(r,\theta) dr d\theta =  r \,dr d\theta  \left[ 1 + \left(\frac{r}{r_0} \right)^{-\gamma} \right]  {\Big /}
\pi R_{\rm out}^2 \left[ 1 + \frac{2}{2-\gamma} \left(\frac{R_{\rm out}}{r_0} \right)^{-\gamma} \right] , \qquad r< R_{\rm out}.
\end{eqnarray}
The first term in square brackets in the numerator represents the contribution for a uniform starfield, while the second term represents the additional contribution due to clustering.  
We use this distribution to compute the averages over object separations by assuming $N \gg 1$  and then taking the limit in which the lensing plane extends over a infinite area  (i.e. so that $R_{\rm out} \gg r_0$),
\begin{eqnarray}
\sigma N \left\langle e^{i {\bf q} \cdot ({\bf r}_k - {\bf r}_l)} \right\rangle 
&=& \sigma N \lim_{R_{\rm out} \rightarrow \infty} \int_0^{R_{\rm out}}  dr \int_0^{2 \pi} d\theta \, e^{i q r \cos \theta} p(r,\theta) \\
&=& -2^{1-\gamma} \gamma \pi \sigma^2  r_0^\gamma q^{\gamma-2} 
\left[ \frac{\Gamma \left(\frac{-\gamma}{2} \right)}{\Gamma \left(\frac{\gamma}{2} \right)} \right] , \qquad \frac{1}{2} < \gamma < 2.
\label{ObjPosnAvg}
 \end{eqnarray}
Evaluation of the integral on the first line above demonstrates that the contribution from the term representing a uniform stellar distribution is identically zero, so that only clustering contributes to this average over object positions.  We also note that the average over object positions is positive since the ratio of the two $\Gamma$ functions in square brackets is negative.

Clustering between dissimilar populations is considered by generalizing the definition of the two point correlation function. The differential probability of finding a star of type 1 in the area $dA_1$ and a star of type 2 in the area $dA_2$ is $dP = \sigma_1 \sigma_2 \left[ 1+ \xi_{12}({\bf r}) \right] dA_1 dA_2$.
The two-point correlation function $\xi_{12}({\bf r})$ can be determined from the individual clustering properties of the two star types.  Motivated by studies of cluster-galaxy clustering in cosmology\footnote{An example of clustering between dissimilar populations in cosmological contexts is the galaxy-cluster two point correlation, which can be compared against the galaxy-galaxy and cluster-cluster two-point correlation functions (see Peebles 1980b; Bahcall 1988).  The galaxy-cluster two-point correlation function is empirically determined to contain two terms, the first of which is important at short scales and is due to the tendency of galaxies to cluster in clusters.  In our treatment we assume that stars of a certain type have no tendency to cluster around stars of a different type, so this term is unimportant.  The second term is related to the clustering of clusters themselves, and is important in our case.  This contribution is given approximately by the geometric mean of the two-point correlation functions for objects of either type: $\xi_{GC} \approx \xi_{CC}^{1/2} \xi_{GG}^{1/2}$.}, the two-point correlation function is approximated by the geometric mean of the two-point correlation functions for objects of either type: $\xi_{kl}({\bf r}) = \xi_{kk} ({\bf r})^{1/2} \xi_{ll}({\bf r})^{1/2}$ for clustering between stars of masses $M_k$ and $M_l$.  With this generalization it is straightforward to calculate the contribution of the cross term as, 
\begin{eqnarray}
\sigma_{M_k}(N_{M_l}-1) \left\langle e^{i {\bf q} \cdot ({\bf r}_{i,k} - {\bf r}_{j,l})} \right\rangle = -2^{1-\gamma} \gamma \pi \sigma_{M_k} \sigma_{M_l} r_{0,M_k}^{\gamma/2} r_{0,M_l}^{\gamma/2} q^{\gamma-2} \frac{\Gamma \left( -\gamma/2 \right)}{\Gamma \left( \gamma/2 \right)}, 
\end{eqnarray}
where $r_{0,M_k}$ and $r_{0,M_l}$ are the clustering scale lengths of stars with masses $M_k$ and $M_l$ respectively.

Although we have only explicitly considered stellar clustering in terms of a power-law two-point correlation function above, the generalization to other two-point correlation functions is obvious.




\section{Power spectrum of intensity fluctuations from stellar lensing} 
Having obtained expressions for the power spectrum of mass density fluctuations across the lensing plane we proceed to evaluate the contribution of stellar microlensing on gravitational radiation. 

\subsection{Lensing by a single population} 
We first investigate the character of the intensity fluctuations due to only a single lensing population of stars.  Inspection of eq.\,(\ref{PwrSum}) shows that there are two distinct contributions to the intensity variance, the first from `self-terms', which is proportional to the stellar number density, $\sigma$, and which we hereafter refer to as the `incoherent' lensing contribution in light of the fact that the term is independent of the spatial distribution of the stars.   The second term, whose contribution is proportional to $\sigma^2$, is labelled the `coherent' lensing contribution in light of the fact that this term only contributes if the stars are clustered.  These two contributions are most conveniently treated separately. When the stars are distributed randomly but uniformly over the lensing plane only the incoherent term  contributes to the intensity variations.  

Intuitively one expects both the clustering properties and the Fresnel scale to strongly influence the nature of the intensity fluctuations.  The lensing properties depend on the mass fluctuations contained within the coherence area determined by the Fresnel scale.  The clustering amplitude determines the area over which the density fluctuation power spectrum scales quadratically with the stellar number density, rather than only linearly.  The ratio of the clustering amplitude to the Fresnel scales thus bears on the character of the intensity fluctuations.  For instance, we might expect that lensing by stars clustered on scales much smaller than the Fresnel scale would be equivalent to lensing by a uniformly distributed starfield, in which the power spectrum of intensity fluctuations scales only linearly with the stellar density.

\subsubsection{The contribution from uniformly distributed stars}

The Fresnel scale for the wavelengths typical of gravitational radiation is far larger than the radii of the individual lensing stars.  Thus one expects the particular profiles of the lensing constituents to make a negligible difference to the magnitude of the intensity variance.  For simplicity we employ the $q < \sqrt{5}\,R^{-1}$ approximation to the power spectrum of stellar mass profile, $|\tilde f_M({\bf q}) |^2 = M^2$, and justify its use at the end of this section by examining the length scales of the fluctuations that  contribute most strongly to the intensity variance.  The intensity variance due to the incoherent term is, 
\begin{eqnarray}
\langle \Delta I^2 \rangle_{\rm ic} = 4 \sigma M^2 K^2 \int d^2{\bf q}  \, q^{-4} \sin^2 \left(\frac{q^2 r_{\rm F}^2}{2} \right) = \pi^2 \sigma M^2 K^2 r_{\rm F}^2. \label{Ivar1}
\end{eqnarray}
The power spectrum of temporal intensity fluctuations associated with the incoherent term is,
\begin{eqnarray}
W_{\rm ic}(\omega) = \frac{8 K^2 M^2 \sigma}{v} \int_0^\infty dq \, \left(\frac{\omega^2}{v^2} + q^2 \right)^{-2} \left[1 - \frac{\left(\frac{\omega^2}{v^2} + q^2 \right) R^2}{5} \right] 
\sin^2 \left[ \frac{\left(\frac{\omega^2}{v^2} + q^2 \right) r_{\rm F}^2}{2}\right], 
\end{eqnarray}
where we have now retained terms to second order in $q$ in the spectrum of the stellar mass profile, $|\tilde f({\bf q})|^2$.  The temporal fluctuation spectrum is evaluated separately in the two regions $q<r_{\rm F}^{-1}$ and $q > r_{\rm F}^{-1}$.  In the first region the sine-squared term in eq. (\ref{Womega}) is linearised, and in the second its argument varies rapidly with $q$, so we approximate it by its mean value of $1/2$.  For $\omega /v < r_{\rm F}^{-1}$ this approximation yields,
\begin{eqnarray}
W_{\rm ic}(\omega) = \frac{2 \pi K^2 M^2 \sigma r_{\rm F}^3}{v}  \, \left(1 - \frac{ R^2 \omega^2}{5 v^2} - \frac{R^2}{15 r_{\rm F}^2} \right). 
\end{eqnarray}
The leading term inside the brackets dominates over the remaining terms by a factor larger than 
$r_{\rm F}^2/R^2 \gg 1$.  In the high-frequency, $\omega/v > r_{\rm F}^{-1}$, regime one has,
\begin{eqnarray}
W(\omega) &=& \frac{2 \pi K^2 M^2 \sigma}{v} 
\left[ \frac{ v^3}{\omega^3}\left(\frac{\pi}{2} - \tan^{-1} \frac{v}{r_{\rm F} \omega} \right) 
- \frac{r_{\rm F} v^4}{v^2 \omega^2 + r_{\rm F}^2 \omega^4} \right], \qquad r_{\rm F}^{-1} < \frac{\omega}{v} < R^{-1} \nonumber \\
&\approx& 2 \pi K^2 M^2 \sigma \, \frac{v^2}{\omega^3} \left(\frac{\pi}{2} - \frac{2 v}{r_{\rm F} \omega} \right), \label{W1}
\end{eqnarray}
where again the leading term inside the brackets dominates.  One thus sees that the temporal power spectrum is constant for $\omega / v \la r_{\rm F}^{-1}$ and then decreases as $\omega^{-3}$ at larger angular frequencies.  Since the contribution to the intensity variance scales as $\omega W(\omega)$, structure with wavenumbers $q \sim r_{\rm F}^{-1}$ dominates the contribution to the intensity fluctuations, and intensity fluctuations occur most prominently on a timescale $\tau \sim r_{\rm F}/v$.  

For the wavelengths typical of gravitational radiation the Fresnel scale far exceeds the radii of the individual lensing stars, this result also justifies our original approach in approximating the stellar mass profile $|\tilde f_M({\bf q})|^2$ to lowest order in $q$ only.   This approximation is also valid for the lensing of radio-wave electromagnetic radiation at cosmological distances.  However, at shorter wavelengths the approximations used above fail, where the radii of the stars themselves exceed the Fresnel scale, the above integrals need to be re-evaluated using a precise expression for the stellar mass profile.

\subsubsection{The additional contribution from clustering}

\begin{figure}
\begin{center}
\begin{tabular}{cc}
\psfig{file=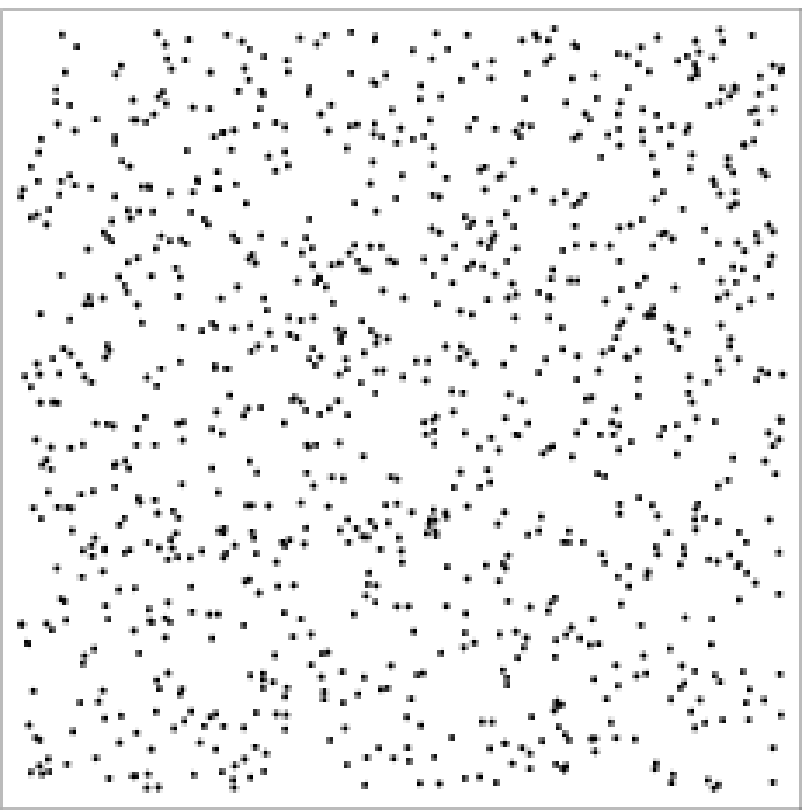,width=6cm} & \psfig{file=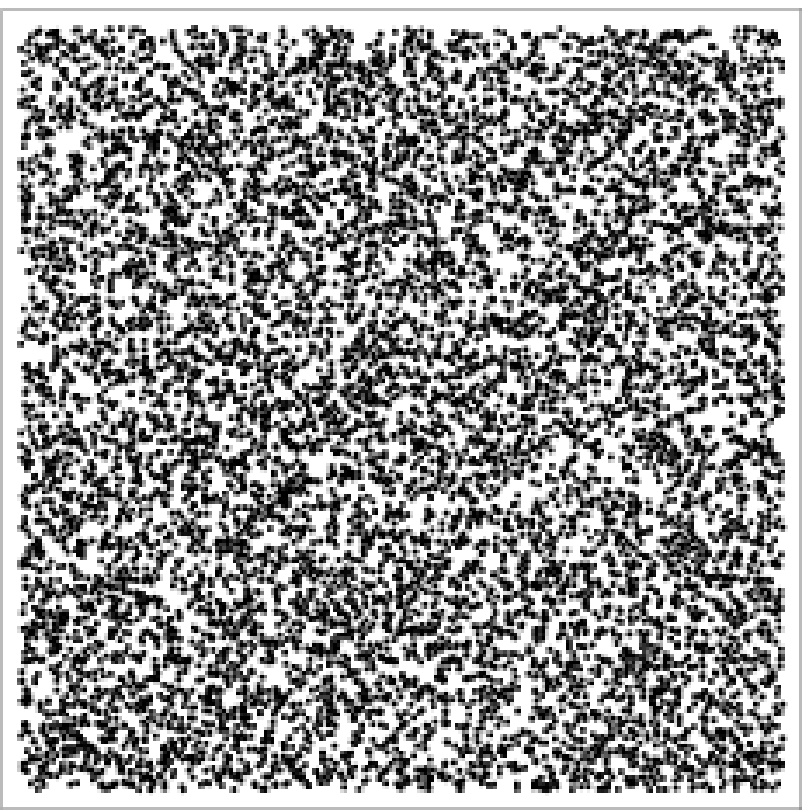,width=6cm} \\
\psfig{file=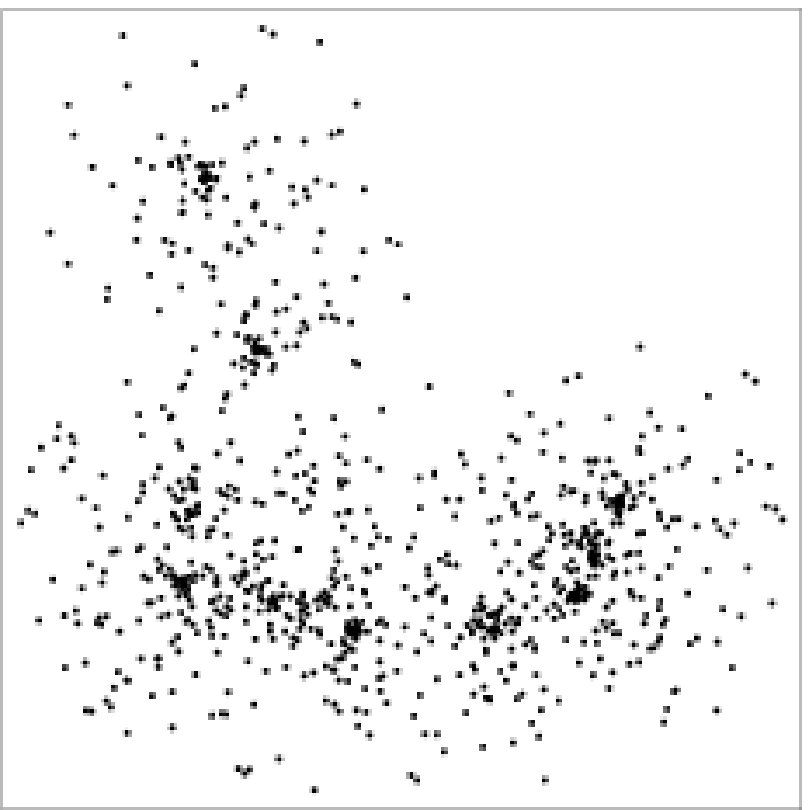,width=6cm} & \psfig{file=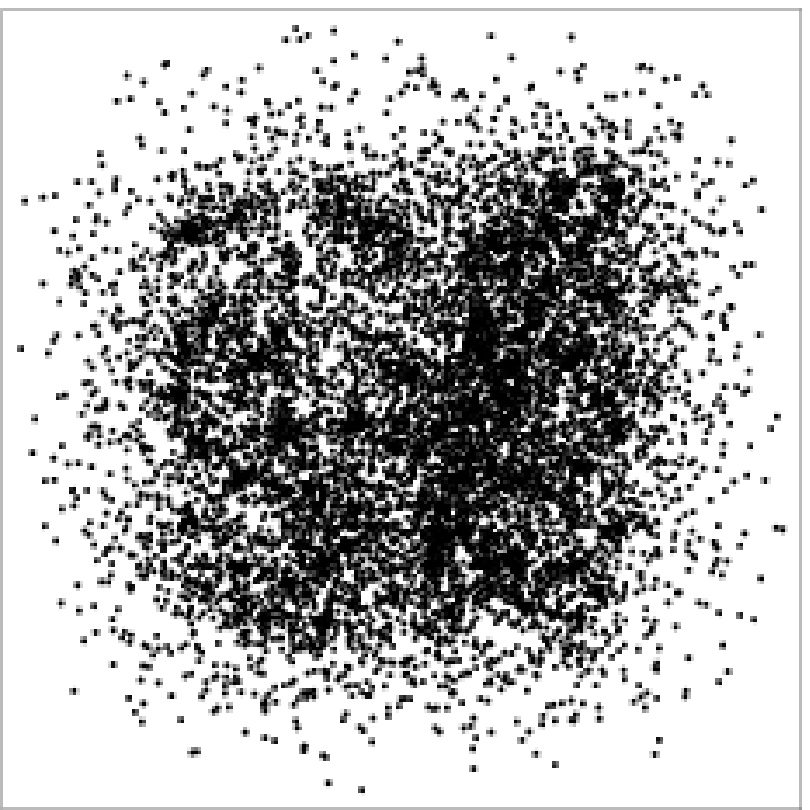,width=6cm} \\
\end{tabular}
\end{center}
\caption{An illustration of the importance of clustering in determining the contribution of stellar microlensing lensing.   The top row depicts a starfield in which the probability distribution of stellar positions is uniform across the lensing plane,  while the bottom row illustrates a distribution in which there is a contribution from clustering.  Each left panel contains $10^4$ stars, and each right panel $10^5$ stars.  As the stellar density increases the mass surface density begins to look relatively uniform for panels on the top row, leading to a uniform phase delay across the lensing plane compared to the case in which clustering is important.} \label{ClustContrib}
\end{figure}

The foregoing results provide a lower bound to the contribution from stellar microlensing.  However, the contribution can be much greater than if the stellar density in the lensing region is sufficiently large and the stars exhibit a tendency to cluster.  Figure~\ref{ClustContrib} demonstrates why one intuitively expects the contribution of clustering to dominate once the stellar density becomes large.  For an unclustered stellar distribution the rms deviation in mass density increases only as the square-root of the density, whereas the contribution from clustered stars increases linearly.   

The contribution to the intensity variance from clustering terms is evaluated using the average over object positions in eq. (\ref{ObjPosnAvg}), and again using the approximation $| \tilde f({\bf q})|^2 = M^2$,
\begin{eqnarray}
\langle \Delta I^2 \rangle_{\rm coh} &=& 8 \pi K^2 \int q\,dq \, q^{-4} \sin^2 \left(\frac{q^2 r_{\rm F}^2}{2} \right) \sigma N \left\langle \exp[i {\bf q} \cdot ({\bf r}_l - {\bf r}_k)] \right\rangle \nonumber \\
&=&  \frac{2^{4-\gamma} \pi^2 \gamma \, \Gamma \left( -\frac{\gamma}{2}\right) }{6\gamma - \gamma^2- 8} \cos \left( \frac{\pi \gamma}{4} \right)
K^2 \sigma^2 M^2  r_{\rm F}^4 \left( \frac{r_0}{r_{\rm F}} \right)^\gamma , \qquad 1/2< \gamma < 2.
\label{Ivar2}
\end{eqnarray}
The importance of clustering in determining the lensing characteristics is illustrated by the fact that this contribution from the cross terms is a factor $\sim \sigma r_{\rm F}^2 \left(r_0/r_{\rm F} \right)^\gamma$ larger than that from the self-terms.   This result is easily understood when one considers that the extra number of stars within the Fresnel radius when clustering is important is $N(r<r_{\rm F}) \sim  \pi r_{\rm F}^2 \sigma (r_{\rm F}/r_0)^{-\gamma}$.  The stellar number density need not be very high for clustering to dominate the amplitude of the intensity fluctuations; one requires a stellar number density that exceeds one star over an area $r_{\rm F}^2 (r_0/r_{\rm F})^{\gamma}$.  This area can be of order many square parsecs if the lensing occurs at cosmological distances.

The temporal power spectrum is again evaluated by approximating the sine-squared function as $q^4 r_{\rm F}^4/4$ for $q < r_{\rm F}^{-1}$ and as 1/2 for $q > r_{\rm F}^{-1}$.  The calculation of the spectrum is detailed in Appendix \ref{Wcalc}, and we merely list the results here:
\begin{eqnarray}
W_{\rm coh} (\omega) &=& \sigma^2 M^2 K^2 r_0^\gamma \pi^2 \gamma  \sin \left( \frac{\pi \gamma}{2} \right)
 \Gamma(-\gamma)
  	\times \left\{ 
\begin{array}{ll}
- 4   \, \frac{ r_{\rm F}^4}{\omega} \left( \frac{\omega}{v} \right)^\gamma, 
&  \frac{\omega}{v} < r_{\rm F}^{-1},\, 1/2 < \gamma < 2.  \\
8 \frac{ (\gamma-3) (\gamma-1)}{(\gamma-2)(\gamma-4)}   \frac{1}{v} \left( \frac{\omega}{v} \right)^{\gamma-5}, &
 r_{\rm F}^{-1} < \frac{\omega}{v} < R^{-1},  \, 1/2< \gamma < 2 . 
\end{array} 
\right. \label{CrossTermPwrLowOmega} \label{W2i}  \label{CrossTermPwrHighOmega} \label{W2f}
\end{eqnarray}
The power spectrum scales as $\omega^{\gamma-1}$ in the low frequency regime but decreases sharply in the regime $\omega > v \,r_{\rm F}^{-1}$.   Most of the fluctuation power, which scales as 
$\omega W(\omega)$, is therefore sharply concentrated on angular frequencies $\omega \sim v r_{\rm F}^{-1}$.


\subsection{Generalization to a distribution of stellar masses}
The generalization to a population of stars requires a specification of both the mass distribution and clustering properties.  The mass distribution of stars is taken to be a power law with index $\beta$, so that if the total stellar number density is $\sigma$ the surface density of stars of masses between $M$ and $M+dM$ is,
\begin{eqnarray}
\sigma_M \, \frac{dM}{M_\odot}= C \sigma \left( \frac{M}{{\rm M}_\odot}\right)^{-\beta} \frac{dM}{M_\odot}, \qquad  M_{\rm min} < M < M_{\rm max},
\end{eqnarray}
between lower and upper cutoffs $M_{\rm min}$ and $M_{\rm max}$ respectively.  The constant of normalization, $C=(1-\beta) [(M_{\rm max}/M_\odot)^{1-\beta} - (M_{\rm min}/M_\odot)^{1-\beta}]^{-1}$, is chosen so that integration of $\sigma_M$ over the mass distribution equals the total stellar number density, $\sigma$.  A practical specific choice is a Salpeter initial mass function (Salpeter 1955), in which $\beta = 2.35$, the mass distribution ranges between 0.1 and 125\,M$_\odot$ and the constant of normalization is $C=0.0603$.

If there are sufficiently many stars one converts to the continuum limit in which the sums over object masses in eq.\,(\ref{PhiCross}) are written as integrals, so the mass power spectrum becomes,
\begin{eqnarray}
\Phi_\Sigma ({\bf q}) &=& C \sigma \int_{M_{\rm min}}^{M_{\rm max}}  \left\vert \tilde f_M({\bf q}) \right\vert^2 \left( \frac{M}{{\rm M}_\odot} \right)^{-\beta} \left[1 - 2^{2-\gamma} \gamma \pi^2  r_0^\gamma(M) q^{\gamma-2} C \sigma \frac{\Gamma(-\gamma/2)}{\Gamma(\gamma/2)}  \left( \frac{M}{{\rm M}_\odot} \right)^{-\beta}  \right] \frac{dM}{M_\odot} \nonumber \\
&\null& \qquad 
- 2^{1-\gamma} \gamma \pi q^{\gamma-2} C^2 \sigma^2 \frac{\Gamma(-\gamma/2)}{\Gamma(\gamma/2)} 
 \iint_{M_{\rm min}}^{M_{\rm max}}  \tilde f_{M_1}({\bf q}) \tilde f_{M_2}^*({\bf q}) r_0^{\gamma/2}(M_1) r_0^{\gamma/2} (M_2) 
\left( \frac{M_1}{{\rm M}_\odot} \right)^{-\beta} \, \left( \frac{M_2}{{\rm M}_\odot} \right)^{-\beta} \frac{dM_1 dM_2}{M_\odot^2}, \nonumber \\
\label{MassDistnSpect}
\end{eqnarray}  
where $r_0$ is parameterized in terms of stellar mass.  The radii of the lensing stars also depend on the mass,  but this only changes the mass power spectrum at wavenumbers $q \sim R^{-1}$.  This wavenumber remains sufficiently large even for the largest stars that it does not appreciably alter the lensing properties because, as shown  above, the dominant contribution to the intensity variance arises from mass fluctuations on length scales $\sim r_{\rm F} \gg R$ for the wavelengths typical of gravitation radiation.  However, for lensing of electromagnetic radiation at wavelengths shorter than $\sim 0.01\,$m this simplifying assumption fails, and the density profiles of the stars must be taken into account; this case is not investigated further here.

\subsubsection{A specific model for the mass and clustering distributions}
We introduce here a simple analytically tractable model which, though possibly unrealistic as a complete description of stellar clustering, is nevertheless useful as a context in which to explore certain properties of clustering germane to an understanding of the underlying physics of stellar microlensing.

Our prescription for the clustering length scale is founded on the basis that long-lived stars diffuse far from their birthplace over their lifetime and thus cluster weakly, but higher mass stars are more short-lived and are thus highly clustered because they have less time to diffuse from their birthplaces, in which they are originally highly clustered.  We thus postulate a clustering amplitude of the form,
\begin{eqnarray}
r_0^\gamma(M) = A \left( \frac{M}{{\rm M}_\odot} \right)^{a}, \label{ClusterFunction}
\end{eqnarray}
where the positivity of $a$ reflects the fact that less massive stars are more uniformly distributed and thus possess smaller clustering amplitudes.
The intensity variance is calculated by replacing the integrals over object masses in expressions (\ref{Ivar1}) and (\ref{Ivar2}) to give,
\begin{eqnarray}
\langle \Delta I^2 \rangle &=& \frac{M_\odot^2 C \sigma \pi^2 K^2 r_{\rm F}^2}{3-\beta} \left[ \left( \frac{M_{\rm max}}{M_\odot} \right)^{3-\beta} - \left( \frac{M_{\rm min}}{M_\odot}\right)^{3-\beta} \right]  
- 2^{4-\gamma} \gamma K^2 \pi^2 \cos \left( \frac{\pi \gamma}{4}\right) r_{\rm F}^{4-\gamma} 
 C^2 \sigma^2 A M_\odot^2 
\frac{\Gamma \left( -\gamma/2 \right)}{8-6\gamma + \gamma^2} \nonumber \\
&\null& \qquad \qquad  \times
\left\{
\frac{1}{3+a -2 \beta} 
\left[ \left( \frac{M_{\rm max}}{M_\odot} \right)^{3+ a  -2 \beta} -\left( \frac{M_{\rm min}}{M_\odot} \right)^{3+ a  -2 \beta} \right]
+ \frac{1}{\left( 2+\frac{a}{2} - \beta \right)^2} 
\left[ \left(\frac{M_{\rm max}}{M_\odot} \right)^{2+\frac{a }{2} - \beta} - \left(\frac{M_{\rm min}}{M_\odot} \right)^{2+\frac{a }{2} - \beta} \right]^2 
\right\}, \nonumber \\
		\label{IvarTotal}
\end{eqnarray}
where the $q< \sqrt{5} R^{-1}$ approximation to $| \tilde f_M({\bf q}) |^2$ is again employed.

Similarly, to obtain the power spectrum of intensity fluctuations one makes the replacement, 
\begin{eqnarray}
\sigma M &\rightarrow& \frac{M_\odot^2 C \sigma}{3-\beta} \left[ \left( \frac{M_{\rm max}}{M_\odot} \right)^{3-\beta} - \left( \frac{M_{\rm max}}{M_\odot}\right)^{3-\beta} \right] ,
\end{eqnarray}
in eq. (\ref{W1}),  and the following replacements in eqs. (\ref{W2i}) -- (\ref{W2f}),
\begin{eqnarray}
\sigma^2 M^2 r_0^\gamma &\rightarrow&   C^2 \sigma^2 A M_\odot^2 \left\{ \frac{1}{ 3+a  - 2 \beta } \left[ \left( \frac{M_{\rm max}}{M_\odot} \right)^{3 + a -2 \beta} -\left( \frac{M_{\rm min}}{M_\odot} \right)^{3+a  -2 \beta} \right]  
+ 
\frac{1}{\left( 2+ \frac{a}{2} - \beta \right)^2} 
\left[ \left(\frac{M_{\rm max}}{M_\odot} \right)^{2+\frac{a }{2} - \beta} - \left(\frac{M_{\rm min}}{M_\odot} \right)^{2+\frac{a}{2}- \beta} \right]^2 \right\}.  \nonumber \\
\end{eqnarray}
  
The clustering contribution from stars of various masses depends on how the clustering depends on mass.  It might be supposed that massive stars with short lifetimes diffuse relatively small distances from their natal molecular clouds, and thus cluster strongly compared to dwarf stars, which may diffuse long distances over their much longer lifetimes.  For a distribution following a Salpeter mass distribution low mass stars make the dominant contribution to the intensity variance when $a< 0.7$, both low- and high-mass  stars are important in the range $0.7 < a < 1.7$, while high-mass stars make the dominant contribution to the intensity fluctuations when $a > 1.7$.

\subsection{Summary of results}

We provide here a brief summary of the important results pertaining to the lensing of gravitational radiation by stars derived in the preceding two sections.  The amplitude of the intensity fluctuations is determined by the mass density power spectrum.  For lensing by discrete objects, this power spectrum contains two sets of terms: (i) a set of self-terms which is linearly proportional to the object density, $\sigma$, and is independent of how the objects are distributed across the lensing plane and, (ii) a set of cross-terms which increases quadratically with the stellar density but is only important if the objects are clustered.   The latter terms are important for gravitational radiation because the stars may cluster on length scales comparable to the coherence radius $r_{\rm F}$, so that a collection of $N$ clustered stars acts like a single lensing macrostar of mass $M N$.  The lensing contribution from the same number of uniformly distributed objects is equivalent to an object of mass $M \sqrt{N}$ only.  In a simple case in which stars are clustered across the lensing plane according to a power-law two-point correlation function $\xi({\bf r}) = (r/r_0)^{-\gamma}$, the contribution to the intensity variance from clustered stars exceeds the uniform component by a factor $\sigma r_{\rm F}^2 (r_0/r_{\rm F})^\gamma$.   

The effect of lensing by objects of different masses is considered by investigating the effect of lensing from objects whose mass distribution follows a power law.  For uniformly distributed stars the intensity variance is dominated by the high mass cutoff of the object distribution provided that the mass distribution falls less steeply than $M^{-3}$.  Thus lensing by a starfield that follows a Salpeter initial mass function (i.e. a distribution proportional to $\sim M^{-2.35}$) is dominated by the high mass cutoff when clustering is negligible. 

The power spectrum of temporal intensity fluctuations is also evaluated for both uniformly-distributed and clustered objects.  It is shown that in both cases the largest contribution to the fluctuations occurs on the timescale required for the objects on the lensing plane to drift a distance $\sim r_{\rm F}$ across the line of sight.  The lensing is also most sensitive to mass fluctuations on this length scale.

\section{Signatures of lensing and magnitude estimates} \label{Discussion}


We present here several simple scaling relations that allow the magnitude of intensity fluctuations to be computed for any given lensing geometry and mass fluctuation amplitude.  We defer to a later a work
a more rigorous analysis that takes into account the distance ranges out to which the various types of source might be observed by current and future detectors.  

The magnitude of the intensity variations depends on the strength of the scattering, which is set by the amplitude of the mass power spectrum relative to the Fresnel scale.  It is convenient to define an effective distance to the scattering material, $D_{\rm eff} =D_L D_{LS}/D_S$ so that the Fresnel scale takes the form $r_{\rm F}=[\lambda D_{\rm eff}/(1+z_L)]^{1/2}$.  When considering scattering by an intervening system this effective lensing distance is of order the angular diameter distance to the system itself.  In terms of values normalised for values typical of scattering by matter from a source at a cosmological distance (e.g. a galaxy), the Fresnel scale for lensing at a distance $D_{\rm eff}$ is, 
\begin{eqnarray}
r_{\rm F} &=& 1.2 \, (1+z_L)^{-1/2} \, \left( \frac{\nu}{1\,{\rm Hz}} \right)^{-1/2} \, \left(\frac{D_{\rm eff}}{1\,{\rm Gpc}} \right)^{1/2} \, {\rm pc}. \label{FresnelNumerical}
\end{eqnarray}

\subsection{Lensing by a power law spectrum of mass fluctuations}

For a power law spectrum of mass fluctuations the scale length of the phase changes that give rise to intensity variations is 
\begin{eqnarray}
r_0 = \left\{ \begin{array}{ll}
 665 \, (1+z_L)^{-2/3}  \left(\frac{\nu}{1\,{\rm Hz}} \right)^{-2/3} \left( \frac{l_0}{1\,{\rm pc}} \right)^{-1/3} 
 \left( \frac{\langle \Delta \Sigma^2 \rangle}{1\,{\rm M}_\odot^2\,{\rm pc}^{-4}} \right)^{-1/3}  {\rm pc},  & \beta=1, \\
213 \, \left( 1+z_L \right)^{-1/2} \left( \frac{\nu}{1\,{\rm Hz}} \right)^{-1/2} \left( \frac{\langle \Delta \Sigma^2 \rangle}{1\,{\rm M}_\odot^2\,{\rm pc}^{-4}}\right)^{1/4}  {\rm pc}, & 2< \beta < 4, \\
\end{array} \right. 
\end{eqnarray}
where, in order to ensure that the numerical constants do not themselves depend on $\beta$, the case $\beta=1$ is chosen to represent the regime $0 < \beta < 2$.  
Geometric phase delays thus dominate phase delays induced by inhomogeneities in the mass distribution, so that the intensity fluctuations occur in the weak perturbation regime.   

In calculating the magnitude of lensing effects on gravitational radiation it is important to recognise that gravitational wave detectors measure the wave strain directly, so it is most appropriate to consider fluctuations in the magnitude of $\phi$, where even a small intensity variance implies potentially measurable fluctuations in the gravitational wave amplitude.
We express the variability in terms of the rms wave amplitude fluctuation, $\langle \delta \vert \phi \vert^2 \rangle^{1/2}$.  In the regime of weak fluctuations it can be shown using the Rytov method that the intensity fluctuations are log-normally distributed (Tatarski 1967; Fante 1968) so that here, where the fluctuations are very small, this is well-approximated by a normal distribution (see Goodman (1985) and references therein).  The rms wave amplitude fluctuation, $\langle \delta \vert \phi \vert^2 \rangle^{1/2}$, is thus expressed in terms of the intensity variance as $(2/\pi)^{1/4} \langle \Delta I^2 \rangle^{1/4}$.  We also explicitly write observational quantities  in this section in terms of the mean wave strain of the source, $\langle | \phi | \rangle$, which is the value that would be measured in the absence of lensing effects.  (In previous sections this normalization was made implicit by assuming the source to be of unit amplitude (cf. Sect.\,2).)  
The rms wave amplitude fluctuation is thus 
\begin{eqnarray}
\langle | \phi |^4 \rangle^{1/4} &=&   0.013 \,(1+z_L)^{(2-\beta)/8} \left( \frac{\nu}{1\,{\rm Hz}} \right)^{1/8} 
\left( \frac{l_0}{1\,{\rm pc}}\right)^{1/4} \left( \frac{\langle \Delta \Sigma^2 \rangle }{1\, {\rm M}_\odot^2\,{\rm pc}^{-4}} \right)^{1/4} 
\left( \frac{D_{\rm eff}}{1\,{\rm Gpc}} \right)^{3/8}, \qquad \beta = 1, \label{IsqrEst1} \\
&=&
0.012 \, \left( \frac{\langle \Delta \Sigma^2 \rangle }{1\, {\rm M}_\odot^2\,{\rm pc}^{-4}} \right)^{1/4} \left( \frac{D_{\rm eff}}{1\,{\rm Gpc}} \right)^{1/2}, \qquad 2 < \beta < 4. \label{IsqrEst2}
\end{eqnarray}
If the wave amplitude fluctuations differ from a Gaussian distribution the results will differ only slightly from those in eqs. (\ref{IsqrEst1}) \& (\ref{IsqrEst2}).  This is because the rms of the wave amplitude fluctuations differs by only a value of order unity to the power of one quarter if the phase fluctuations deviate from a normal distribution.  We note that the estimates in eqns. (\ref{IsqrEst1}) \& (\ref{IsqrEst2}) are similar to those computed by Takahashi (2006), who considered fluctuations in the magnitude and phase of $\phi$ directly.

The magnitude of lensing-induced variations depends on the mass surface density variance.  This quantity may be very large if the line of sight intersects a massive structure, such as a galaxy cluster.  We therefore compute a worst-case upper limit to this quantity based on measurements of the mass surface density in galaxy cluster environments using weak lensing measurements.  A suitable fiducial mass density $\Sigma \approx  140 \,{\rm M}_\odot\,{\rm pc}^{-2}$ is derived from weak lensing measurements of the cluster Cl 1358$+$62 (Hoekstra et al. 1998).  If we assume that the bulk of this matter is distributed inhomogeneously over the extent of the cluster one has $\langle \Delta \Sigma^2 \rangle \sim 2.0 \times 10^4 \, {\rm M}_\odot^2\,{\rm pc}^{-4}$.  The rms wave amplitude fluctuation is thus $\approx 15\,$\% for the fiducial numbers used in the scaling relations (\ref{IsqrEst1}) \& (\ref{IsqrEst2}) above.




\subsection{Lensing by stars}

An estimate of the stellar density on the lensing plane is crucial in estimating the magnitude of the stellar lensing effects.  The density also determines the number of stars that contribute to the lensing at any one instant.  The number of stars that contribute simultaneously to the lensing is determined by the coherence area,  whose radius is comparable to the Fresnel scale, which depends on the effective lensing distance.  For lensing within our Galaxy one has $D_{\rm eff} \sim \Delta L/2 \sim 5\,$kpc for a medium of thickness $\Delta L\sim 10\,$kpc, whereas for lensing of an object at cosmological distances by an intervening system, the fiducial effective lensing distance is of order $1\,$Gpc.  For lensing within the Galaxy the number of stars that contributes to the lensing is, 
\begin{eqnarray}
N_{\rm MW} \sim 0.2 \nu^{-1} \left(\frac{\rho_\star}{1\,{\rm pc}^{-3}} \right) \left(\frac{\Delta L}{ 10\,{\rm kpc}} \right)^2,
\end{eqnarray}
for a stellar volume density $\rho_\star$.  The number of stars that might contribute to lensing by an intervening system for a source located at cosmological distances is much greater,
\begin{eqnarray}
N_{\star \rm IG} \sim 10^4 (1+z_L)^{-1} \nu^{-1} \left(\frac{\rho_\star}{1\,{\rm pc}^{-3}} \right) \left(\frac{\Delta L}{ 1\,{\rm kpc}} \right) \left( \frac{D_{\rm eff}}{1\,{\rm Gpc}} \right).
\end{eqnarray}
Thus the statistical approach adopted here is nearly always applicable to intergalactic scattering, but is only useful for Galactic lensing at frequencies $\nu \la 10^{-2}\,$Hz.  The following discussion is restricted to regimes in which there are sufficiently many lensing objects that a statistical approach is useful.  The effect of lensing by only a few discrete objects has been considered extensively elsewhere (e.g. Chang \& Refsdal 1979, 1984; Kayser, Refsdal \& Stabell 1986; Schneider \& Weiss 1987; Wambsganss 1990; Lewis et al. 1993).

The simplest feasible lensing scenario consists of lensing by a large number of identical stars.  This approach presents a serviceable description of lensing by dwarf stars whose spatial distribution is likely to be devoid of clustering because their long lifetimes allow them to diffuse homogeneously throughout a galaxy.  The contribution from such stars is described by eq. (\ref{Ivar1}), which gives,
\begin{eqnarray}
\frac{\langle \delta \vert \phi \vert^2 \rangle_{\rm ic}^{1/2}}{\langle |\phi| \rangle} 
= 0.019 \, (1+z_L)^{1/4} \left( \frac{\nu}{1\,{\rm Hz}} \right)^{1/4} 
\left( \frac{D_{\rm eff}}{1\,{\rm Gpc}}\right)^{1/4} \left( \frac{\sigma}{100\,{\rm stars\,pc}^{-2} }\right)^{1/4} 
\left( \frac{M}{0.1\,{\rm M}_\odot} \right)^{1/2} .
\end{eqnarray}
The same lensing contribution is obtained if the lensing plane were instead populated by $1\,{\rm M}_\odot$ stars with surface density 1\,star\,pc$^{-2}$.   The generalization of this result to lensing by a distribution of stars of various masses shows that the low mass end of the distribution dominates only if the index of the power-law distribution falls more steeply than $M^{-3}$.  Most expected mass distributions are shallower than this, so the dominant contribution to the lensing amplitude arises from the upper cutoff in the stellar mass distribution.  Taking the distribution to follow a Salpeter initial mass function with $0.1 {\rm M}_\odot < M<125 {\rm M}_\odot$ and $\beta=2.35$, eq. (\ref{IvarTotal}) shows the rms amplitude fluctuation is,
\begin{eqnarray}
\frac{\langle \delta \vert \phi \vert^2 \rangle_{\rm ic}^{1/2}}{\langle |\phi| \rangle} 
= 0.075 \, (1+z_L)^{1/4} \left( \frac{\nu}{1\,{\rm Hz}} \right)^{1/4} 
\left( \frac{D_{\rm eff}}{1\,{\rm Gpc}}\right)^{1/4} \left( \frac{\sigma}{100\,{\rm stars\,pc}^{-2} }\right)^{1/4}.  
\end{eqnarray}

When the stars are clustered there is an additional contribution to the intensity variance.   The magnitude of this contribution depends on the clustering amplitude, which is poorly constrained {\it a priori}.  It is convenient to parameterise the clustering amplitude in terms of an over-density parameter, $\chi$,  which quantizes the factor by which the stellar density exceeds the mean (uniform density) over some fixed separation.   In this way, the number of objects within a region of radius $X$ is $N(r<X) = (1+\chi) \sigma \pi X^2$.  We choose a fiducial distance $X=1\,$pc and accordingly label the overdensity parameter $\chi_{\rm 1pc}$.
The additional clustering contribution to the intensity variance for a single clustered lensing population is thus, from eq. (\ref{Ivar2}),
\begin{eqnarray}
\frac{\langle \delta \vert \phi \vert^2 \rangle_{\rm coh}^{1/2}}{\langle |\phi| \rangle} = 
0.07 \, (1+z_L)^{\gamma/8} \left( \frac{\nu}{1\,{\rm Hz}} \right)^{\gamma/8} 
\left( \frac{D_{\rm eff}}{1\,{\rm Gpc}}\right)^{(4-\gamma)/8} 
\left( \frac{\sigma}{100\,{\rm stars\,pc}^{-2} }\right)^{1/2} 
\left( \frac{M}{0.1\,{\rm M}_\odot} \right)^{1/2} 
\left( \frac{\chi_{\rm 1pc}}{1} \right)^{1/4} 
, \nonumber \\ \label{ClustSingleEst}
\end{eqnarray}
where the numerical constant reflects the choice $\gamma=1$ (cf. eq. (\ref{xi})).  The rms wave amplitude for other values of $\gamma$ is obtained by noting that the amplitude varies slowly, between $0.107$ and $0.128$ for $\gamma$ between $1/2$ and $3/2$ for the fiducial parameters used in eq. (\ref{ClustSingleEst}).   The generalization of this estimate to a distribution of masses requires assumptions about the strength of clustering as a function of mass.  Taking $a=1$ as the index which characterizes the dependence of the clustering properties on mass (cf. eq. (\ref{ClusterFunction})), it remains only to determine the clustering amplitude for a stars of a single mass. If the number overdensity is $\chi_{\rm 1pc}$ for stars of mass $M_{\rm c}$, eq. (\ref{IvarTotal}) shows that the rms wave amplitude is,
\begin{eqnarray}
\frac{\langle \delta \vert \phi \vert^2 \rangle_{\rm coh}^{1/2}}{\langle |\phi| \rangle} = 
0.11 \, (1+z_L)^{1/8} \left( \frac{\nu}{1\,{\rm Hz}} \right)^{1/8} 
\left( \frac{D_{\rm eff}}{1\,{\rm Gpc}}\right)^{3/8} 
\left( \frac{\sigma}{100\,{\rm stars\,pc}^{-2} }\right)^{1/2} 
\left( \frac{\chi_{\rm 1pc}}{1} \right)^{1/4} 
\left( \frac{M_{\rm c}}{10\,{\rm M}_\odot}\right)^{-1/4}, \nonumber \\
\end{eqnarray}
where $\gamma=1$ is assumed.

In the foregoing discussion the contributions from uniformly distributed and clustered stars are treated separately.  The contributions from both are summed to deduce the total intensity variance.  Similarly, the total rms fluctuation in the wave amplitude is determined by summing the incoherent and coherent parts raised to the fourth power (i.e. $\langle \delta \vert \phi \vert^2 \rangle^{2} = (\langle \delta \vert \phi \vert^2 \rangle_{\rm ic}^{1/2})^4 +(\langle \delta \vert \phi \vert^2 \rangle_{\rm coh}^{1/2})^4$).  The total rms is thus well approximated by the larger of the two individual contributions. 

Some remarks on the applicability of our formalism to the lensing of electromagnetic radiation are in order.
To lowest order in the wave amplitude the propagation of both electromagnetic and gravitational radiation are subject to the same wave equation.  The main distinguishing feature between electromagnetic and gravitational radiation is the wavelength and hence the magnitude of the Fresnel scale.  Thus the coherence area and hence the number of stars that contribute to the lensing simultaneously is smaller for electromagnetic radiation.  The smaller value of the Fresnel scale renders the effects of stellar clustering much less important for electromagnetic radiation.  The contribution to the intensity variance from clustering scales as $\lambda^{1-\gamma/2}$.

However, the presence of fewer stars within the coherence area does not invalidate the statistical approach adopted here.  This is because the ensemble-average quantities apply equally whether one averages over space or time.   Obviously a statistical treatment is unnecessary if only a single star contributes to the lensing at a time, but the ensemble average values calculated here are nonetheless correct provided only that the average is of sufficient duration to encompass many lensing objects drifting past the line of sight.    Although the formalism developed here is applicable to electromagnetic lensing, we caution that the final results for intensity variances and power spectra make the approximation that the Fresnel scale greatly exceeds the radii of the lensing objects and thus that the shapes of the individual lensing objects are unimportant.  This is obviously an excellent approximation for gravitational radiation.  However, this approximation fails at optical wavelengths for lensing at cosmological distances.  More generally, the mass profiles of the individual stars must be taken into account for lensing at an effective distance $D_{\rm eff}$ at wavelengths $\lambda \la 100\,(D_{\rm eff}/1\,{\rm Gpc})\,$nm.  When this assumption fails the expressions calculated here for the intensity variance and temporal power spectrum must be recomputed, taking into account the particular mass surface density profiles of the lensing objects (cf. eq. (\ref{fM})).

\subsection{Signatures and timescales of lensing}

Although there are several signatures which could potentially identify radiation that has been lensed by small scale structure, several mitigating factors make their unambiguous detection problematic.  The most obvious characteristics are that the fluctuations are stochastic in nature and that the power spectrum of the variations contains a number of prominent features related to the wavelength and lensing geometry.  The temporal power spectrum exhibits breaks at several characteristic frequencies, most notably at $\omega = v r_{\rm F}^{-1}$, beyond which it steepens.  Analysis of the lensing by both stars and a power law distribution of material shows that the intensity variance is dominated by fluctuations on a timescale $\tau \approx r_{\rm F}/v$:
\begin{eqnarray}
\tau \approx 1.2 \times 10^{3} \, (1+z_L)^{-1/2} \, \left( \frac{\nu}{1\,{\rm Hz}} \right)^{-1/2} \, \left(\frac{D_{\rm eff}}{1\,{\rm Gpc}} \right)^{1/2} \, \left( \frac{v}{1000\,{\rm km\,s}^{-1}} \right)^{-1} {\rm yr}
\end{eqnarray}  
It is difficult to detect temporal fluctuations because most of the relevant fluctuation timescales are too long to be detected.   For a source or lens traversing the line of sight at even $\sim 10^3\,$km\,s$^{-1}$, the expected fluctuation timescale is $> 10^3\,$yr for the estimate of the Fresnel scale in eq. (\ref{FresnelNumerical}) at a frequency of $1\,$Hz.  Shorter timescales are, of course, possible if the source moves at relativistic speeds, but this is unlikely to occur in practice.  The implication here is that any source lensed by small scale structure will not vary appreciably due to lensing, despite the fact that its
intensity is strongly altered by lensing.  This presents a problem for gravitational
wave detectors hoping to remove the effects of stochastic lensing, because the detectors will not be able to average over multiple fluctuations and recover the intrinsic source intensity.

The frequency dependence of the intensity fluctuations depends on the slope of the underlying mass density power spectrum.  For steep, $\beta > 2$, power law spectra the mean intensity variance is independent of frequency, while for shallower spectra the intensity variance exhibits a moderate, $\nu^{1-\beta/2}$, dependence.  The frequency dependence of the intensity variance for lensing by stars is in the range $\nu^{1/4}$ to $\nu^{3/4}$.  Any frequency dependence is potentially observable in chirped (e.g. inspiralling) gravitational wave sources.  The amplitude of the stochastic intensity fluctuations would decrease with time, as the emission frequency increases.  However, identification of the frequency dependence due to lensing is complicated by the fact that the scaling only applies to the ensemble-average intensity variance, whose estimation requires intensity measurements over several fluctuation timescales at each frequency.  This appears to render practical measurements of the frequency dependence unattainable.

\section{Conclusions}

A formalism for treating lensing by (i) objects distributed according to a power law spectrum, and (ii) stars and other discrete objects is considered in the long wavelength limit applicable to gravitational radiation.  
For the large wavelengths typical of gravitational radiation and the large distances characteristic of lensing at cosmological distances the coherence area, of radius $\sim r_{\rm F}$, over which objects influence the observed wave properties exceeds many square parsecs.  Since this area is likely to contain a large number of objects for any line of sight which intersects a galaxy on its way to Earth, a statistical treatment is necessary to compute the effect of microlensing.

For a given amplitude of mass density fluctuations the amplitude of the wave strain increases with the effective lensing distance as $\sim D_{\rm eff}^{1/2}$.  Thus the effect is most pronounced for sources at cosmological distances in which the line of sight intersects a prominent collection of mass, such as a galaxy or proto-galaxy.   The effect is thus much smaller for the lensing of nearby, $D_{\rm eff} \la 100\,$Mpc, sources for two reasons.  Namely, the Fresnel scale (for which $r_{\rm F} \propto D_{\rm eff}^{1/2})$ is small, and the radiation is less likely to encounter an intervening galaxy.  In the absence of lensing by intervening systems only material within the source's own galaxy is likely to contribute to the lensing, for which the effective lensing distance is only of order the size of the galaxy (i.e. tens of kpc at most).

For lensing at cosmological distances stellar microlensing induces rms wave strain fluctuations of order $1-2$\% for unclustered stars and $5-10$\% for clustered stars with stellar densities on the lensing plane of $\sim 100$\,pc$^{-2}$.   However, it may be expected that dark matter, which may also be highly clustered, and whose contribution to the total mass of a galaxy exceeds that of luminous matter, may make a much stronger contribution to the intensity fluctuations.   Lensing by intervening small-scale fluctuations in the dark matter can induce $2-15$\% rms variations in the amplitude of gravitational radiation.

Temporal intensity variations caused by movement of the lensing material across the source-observer line of sight occur on too long a timescale to be observable.  Most fluctuation power is associated with mass variations on scales comparable to the Fresnel scale, and most of the fluctuation power is thus on timescales exceeding $r_{\rm F}/v$, for a transverse speed $v$ plausibly at most $10^3$km\,s$^{-1}$.  The Fresnel crossing time for a lensing speed of $10^3\,$km\,s$^{-1}$ exceeds $1000\,\nu^{-1/2} $years. 

Intensity variations due to lensing provide the opportunity of exploring the nature of mass fluctuations in lensing systems over a large range of scales.  Lensing of $\sim 1\,$Hz gravitational radiation at cosmological distances is sensitive to mass fluctuations on scales of order a parsec.  Radiation detected by LIGO at frequencies $\sim 10\,$kHz is most sensitive to structure on $0.01\,$pc scales, and radiation detected at the bottom end of the LISA band, $\nu \sim 10^{-5}$\,Hz is sensitive to structure on $\sim 300\,$pc scales.  The fluctuation power is attenuated if the lensing medium contains no power on these scales.  Thus investigation of the gravitation wave lensing constitutes a sensitive probe of the dark matter power spectrum over a range of relatively small scales.  
	
We have also considered variations in the mutual coherence of the wave strain, the quantity $\langle \phi({\bf r}') \phi^*({\bf r}'+{\bf r}) \rangle$, which is of interest because it is the most basic estimator of the effect of gravitational lensing on the wave strain.  The degree to which the mutual coherence is a useful statistic is estimated by calculating the variance of this quantity. We find that the amplitude of fluctuations in the mutual coherence is comparable to that expected for intensity fluctuations except when the mutual coherence is measured on baselines $r$ large compared to $r_0$, the scale over which the fluctuations in the phase curvature occur.  This scale is very large so that, in practice, the mutual coherence fluctuates in the same manner as the intensity, $\phi \phi^*$.

\begin{acknowledgements}  
The author thanks Don Melrose and Nissim Kanekar for comments.
The National Radio Astronomy Observatory is a facility of the National Science Foundation operated under cooperative agreement by Associated Universities, Inc.
\end{acknowledgements}

\appendix

\section{The Structure Function for a Power Law Spectrum} \label{DsAppendix}
The phase structure function corresponding to this spectrum is 
\begin{eqnarray}
D_\psi ({\bf r}) &=& 2 K^2 \int d^2{\bf q} \, q^{-4} \Phi_\Sigma ({\bf q}) \left[1 - e^{i {\bf q} \cdot {\bf r}} \right] \nonumber \\
&=& 4 \pi Q_0 K^2 \left\{
\frac{ q_0^2\,q_{\rm min}^{-\beta} r^4}{128} \null_2F_3 \left(1,1;2,3,3; - \frac{q_0^2 r^2 }{4} \right)  
- \frac{q_{\rm min}^{2-\beta} r^4}{128} \null_2F_3 \left(1,1;2,3,3; - \frac{q_{\rm min}^2 r^2 }{4} \right)  + \frac{q_{\rm min}^{-\beta} r^2}{4} \log \left( \frac{q_{\rm min}}{q_0} \right) \right. \nonumber \\
&\null& \left. \qquad \qquad \qquad \qquad
 - \frac{r^{2+\beta} \Gamma \left(-1-\frac{\beta}{2} \right)}{2^{3+\beta} \Gamma \left( 2+\frac{\beta}{2} \right) }  + \frac{q_{\rm min}^{-2-\beta}}{2+\beta} \left[1 - \null_1 F_2 \left(-1-\frac{\beta}{2};1,-\frac{\beta}{2}; - \frac{q_{\rm min}^2 r^2}{4} \right)  \right] \right\} .
\end{eqnarray}
We expand the hypergeometric functions for $q_{\rm min} r \ll 1$, using
\begin{eqnarray}
\null_1 F_2 \left(a_1;b_1,b_2; z  \right) &=& 1+ \frac{a_1}{b_1 b_2}z + {\cal O}(z^2) \label{1F2approx} \\
\null_2 F_3 \left(a_1,a_2; b_1,b_2,b_3;z \right) &=& 1+ \frac{a_1 a_2}{b_1 b_2 b_3}z + {\cal O}(z^2).
\end{eqnarray}
\begin{eqnarray}
D_\psi ({\bf r}) = 4 \pi Q_0 K^2 \left\{ r^2 \frac{q_{\rm min}^{-\beta}}{4} \left[\log \left( \frac{q_{\rm min}}{q_0} \right) - \frac{1}{\beta} \right]  
- r^{2+\beta} \frac{\Gamma \left(-1 - \frac{\beta}{2} \right)}{2^{3+\beta} \Gamma \left( 2+ \frac{\beta}{2} \right)} 
+ r^4 \frac{q_{\rm min}^{-\beta}}{128 (\beta-2)} \left[ q_0^2 (\beta-2) - q_{\rm min}^2 \beta \right] + {\cal O}\left( r^6 q_{\rm min}^{4-\beta} \right) 
\right\}. \nonumber \\
\end{eqnarray}

\section{Characteristics of lensing by a power law mass spectrum} \label{AppC}
\subsection{Power spectrum of intensity fluctuations}

The power spectrum of temporal intensity fluctuations provides information on the relative contributions of various scale lengths to the intensity variance.  The temporal power spectrum is given by,
\begin{eqnarray}
W_I (\omega) = \frac{8 \pi K^2 Q_0 }{v} \left[ 
q_{\rm min}^{-\beta} \int_0^{ \sqrt{{\rm max}(0,q_{\rm min}^2 - \omega^2/v^2)}} dq_y\, 
\frac{\sin^2\left[ \left( \frac{\omega^2}{v^2} + q_y^2\right) \frac{r_{\rm F}^2}{2}\right] }{\left( \frac{\omega^2}{v^2} + q_y^2 \right)^2 } + 
\int_{ \sqrt{{\rm max}(0,q_{\rm min}^2 - \omega^2/v^2)}}^{\infty} dq_y \, 
\frac{\sin^2\left[ \left( \frac{\omega^2}{v^2} + q_y^2\right) \frac{r_{\rm F}^2}{2}\right] }{\left( \frac{\omega^2}{v^2} + q_y^2 \right)^{2+\beta/2} }
\right] . \nonumber \\ \label{WIfull}
\end{eqnarray}
There are three regimes over which the behaviour of the power spectrum changes character for the case $q_{\rm min} < r_{\rm F}^{-1}$: (i) $\omega/v < q_{\rm min} < r_{\rm F}^{-1} $, (ii) $q_{\rm min}<\omega/v < r_{\rm F}^{-1}$ and (iii) $\omega/v > r_{\rm F}^{-1}$.  If instead Fresnel scale exceeds the outer scale ($q_{\rm min} > r_{\rm F}^{-1}$) then only cases (ii), $\omega/v < r_{\rm F}^{-1}$, and (iii) are applicable.

The limiting behaviour in case (i) is evaluated by linearising the sine functions appearing in the integrals.  For cases (ii) and (iii) only the second integral in eq. (\ref{WIfull}) contributes to the power spectrum.  For case (ii) we again linearise the sine function, while in case (iii) the argument of the sine function is much larger than unity and we approximate it by its average value of $1/2$.  This gives the following asymptotic behaviour for $\beta \ga 3/2$:
\begin{eqnarray}
W_I(\omega) \approx \frac{4 \pi K^2 Q_0}{v}
\left\{  \begin{array}{ll} 
q_{\rm min}^{1-\beta} r_{\rm F}^4 \left( 1 + \frac{1}{\beta-1} \right), 
	& \frac{\omega}{v} \la q_{\rm min} \\
\sqrt{\pi} r_{\rm F}^4 \left( \frac{\omega}{v} \right)^{1-\beta} \frac{ \Gamma \left( \frac{\beta-1}{2} \right) }{2 \Gamma \left(\frac{\beta}{2} \right)},
	&  q_{\rm min} \la \frac{\omega}{v} \la r_{\rm F}^{-1} \\
\sqrt{\pi} \left( \frac{\omega}{v} \right)^{-3-\beta}  \frac{\Gamma \left( \frac{3+\beta}{2} \right)
}{\Gamma \left(2 + \frac{\beta}{2} \right)}, 
	& \frac{\omega}{v} \ga r_{\rm F}^{-1}.
\end{array}
\right. \label{WIasymptote}
\end{eqnarray}
Thus, for a given Fresnel scale, we see that the power spectrum consists of a flat portion at low frequencies, a portion over which the spectrum decreases as $\omega^{1-\beta}$ at intermediate frequencies, and a portion that increases even more steeply, as $\omega^{-3-\beta}$, at high frequencies.  In Figure 2 we plot both a numerical integration of eq.\,(\ref{WIfull}) and its asymptotic behaviour as derived in eq.\,(\ref{WIasymptote}).  

The part of the power spectrum that corresponds to short timescales, $\omega/v > r_{\rm F}^{-1}$, contains an oscillatory component that varies approximately as $\sin^2 \omega^2 r_{\rm F}^2/v^2$ and which is not modelled by the asymptotic solution.  The oscillations become increasingly rapid at higher frequencies.  These oscillations are an inherently wave-optics effect attributable to constructive and destructive interference caused by oscillations in the degree of coherence due to the geometric phase delay term (i.e. the first term in the exponential in eq. (\ref{uGrav})).  They may be understood by recalling from elementary optics that scattering from a plane in which the lensing-induced fluctuations are small compared to the geometric phase delay shows a number of concentric bright and dark rings.  This situation is analogous to the familiar pattern displayed in Newton's rings, in which a convex interface placed on a flat surface and illuminated with monochromatic light displays a bright central image surrounded by a number of concentric rings. The dark and bright rings are due respectively to destructive and constructive interference between radiation reflected from the upper and lower surfaces. The central bright peak is due to structure within the first Fresnel zone, the second peak is related to the second ring of coherence surrounding the phase centre, and so on.
Thus we see that the first bump observed in the power spectrum at $\omega/v \sim r_{\rm F}^{-1}$ is due to interference within the first Fresnel zone of the scattered image.  The next peak in the power spectrum is due to interference caused by the second Fresnel zone, and so on.   The period of the Fresnel oscillations in the power spectral domain depends only on the Fresnel scale and the effective lensing velocity.

The most important feature of the intensity power spectrum, however, is that most of the fluctuation power occurs on long timescales.  The contribution to the intensity variance over an interval $\Delta \omega$ scales as $\sim \omega W_I(\omega) \Delta \omega$.  Thus the peak contribution for a mass spectrum steeper than $\beta=2$ occurs at $\omega \sim v q_{\rm min}$.  The contribution over the range $\omega/v = [0,q_{\rm min}]$ exceeds that over the interval $[q_{\rm min},r_{\rm F}^{-1}]$.  This implies that an observation over a time interval $\tau > L_0/v$ is required to obtain an average source intensity close to its ensemble-average (intrinsic) intensity.  If the lensing-induced intensity variance is large, this means that  no practical measurement of the average intensity will approach its ensemble average value for the mass spectrum outer scales and effective lensing speeds likely to be encountered in reality.  The situation is less dire for shallower, $\beta < 2$, spectra, where the peak contribution to the intensity fluctuations occurs on shorter timescales  $\tau \sim r_{\rm F}/v$.    
The divergence of observational quantities from their ensemble-average values is explored further in the discussion below, since it bears ramifications for the suitability of gravitational wave sources as standard sirens.

\begin{figure}
\centerline{\psfig{file=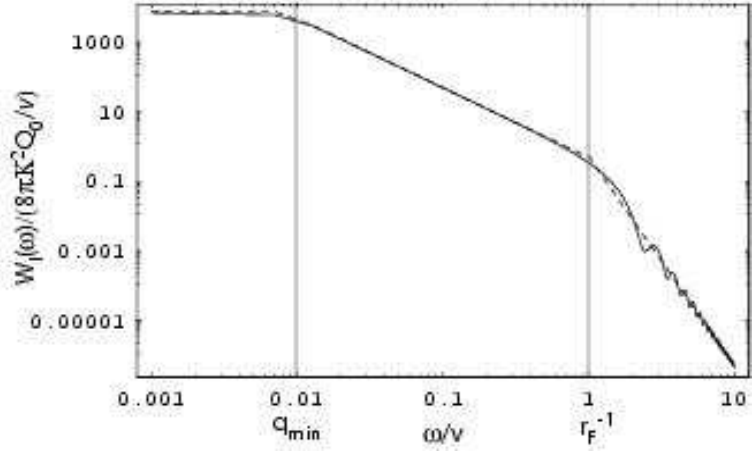,width=10cm}}
\caption{An illustration of the behaviour of the power spectrum of temporal intensity fluctuations, $W_I(\omega)$, for lensing through a power law spectrum of mass fluctuations.  The solid line is obtained by a numerical integration of eq.\,(\ref{WIfull}) with inner scale $q_{\rm min}=0.01$, Fresnel scale $r_{\rm F}=1$ and index $\beta=3$.  The dashed line is the asymptotic solution of eq.\,(\ref{WIasymptote}). }
\end{figure}

\subsection{Fluctuations in the mutual coherence}
The variance of the mutual coherence is evaluated assuming $r_2 < r_{\rm F}$, which allows us to expand the sine-squared function in the integral of eq. (\ref{VisVar}) to second order in ${\bf q} \cdot {\bf r}_2$.  The variance in the visibility measured on a baseline ${\bf r}_2$ is,
\begin{eqnarray}
\langle \Delta V^2({\bf r}_2) \rangle = 4 K^2 \int d^2{\bf q} \, q^{-4} \Phi({\bf q}) \left\{ \sin^2 \left( \frac{q^2 r_{\rm F}^2}{2} \right) -  \frac{{\bf r}_2 \cdot {\bf q}}{2} \sin \left( q^2 r_{\rm F}^2 \right) +  \frac{q^2 r_2^2}{4} \left[1 - 2 \sin^2 \left(\frac{q^2 r_{\rm F}^2}{2} \right) \right] \right\}.
\end{eqnarray}
The first term is recognised as  the intensity variance, while the second term appearing inside the integral is identically zero for any isotropic mass power spectrum.  For the power law mass distribution considered above the third term is evaluated and expanded to lowest order in $q_{\rm min}$ to give
\begin{eqnarray}
\langle \Delta V^2({\bf r}_2) \rangle = \langle \Delta I^2 \rangle - 2 \pi K^2 Q_0 r_2^2 \left[ q_{\rm min}^{-\beta} \log \left( \frac{q_0}{q_{\rm min}} \right) - \frac{q_{\rm min}^{-\beta}}{\beta} \log q_{\rm min}  - \frac{r_{\rm F}^\beta}{2} \cos \left( \frac{\pi \beta}{4} \right) \Gamma \left( -\frac{\beta}{2}\right) \right], \qquad  r_2 < r_{\rm F}. \nonumber \\
\end{eqnarray}

We calculate the power spectrum of fluctuations in the mutual coherence to estimate the timescale on which it fluctuates, and ascertain whether a measurement of the mutual coherence is likely to correspond to its ensemble average value.  The investigation is restricted to the case ${\bf r}_2 \parallel {\bf v}$ since this is the most relevant to observations with a single detector.  The detector measures $V({\bf r}_2= {\bf v} t)$ by measuring the scattered wavefield as it drifts past the detector.  Thus one is most interested in fluctuations in the mutual coherence when the baseline over which it is measured is also oriented parallel to the relative lensing velocity.  The power spectrum of mutual coherence fluctuations is evaluated in the same manner as the intensity power spectrum to find,
\begin{eqnarray}
W_{\Delta V} (\omega; {\bf r}_2)  = \left\{ 
\begin{array}{ll}
P_t (\omega) - \frac{8 \pi K^2 Q_0}{v} r_2 r_{\rm F}^2 q_{\rm min}^{-\beta} \left[ \frac{\pi }{2} + \frac{q_{\rm min}^{-1}}{1+\beta} \left( \frac{\omega}{v} \right) \right] , & \frac{\omega}{v} \la q_{\rm min} \\
P_t(\omega) -  \frac{8 \pi^{3/2} K^2 Q_0 r_2 r_{\rm F}^2}{v} 
\left( \frac{\omega}{v} \right)^{-\beta} \frac{\Gamma \left( \frac{1+\beta}{2} \right)}{\beta \, \Gamma \left( \frac{\beta}{2} \right) },  & q_{\rm min} \la \frac{\omega}{v} \la r_{\rm F}^{-1} \\
P_t(\omega) + {\cal O}\left( r_2^2 \right), & \frac{\omega}{v} \ga r_{\rm F}^{-1}.
\end{array}
\right. 
\end{eqnarray}
Once again, we see that most of the fluctuation power is on the largest scales.  This implies that an observation over a time interval $\tau > L_0/v$ is required to obtain an average visibility close to its ensemble average value for $\beta>2$ spectra, while integration times exceeding $\tau \sim r_{\rm F} / v$ are required when $0 < \beta < 2$.

\subsection{Strong phase perturbations} \label{AppCstrong}
Fluctuations in the strong scintillation regime are evaluated using the phase structure function, eq. (\ref{Ds}), rather than the power spectrum directly.  Strong scattering only applies when $K \, Z({\bf r}_1,{\bf r}_2) > 1$ over the effective region of integration in ${\bf r}_1$ and ${\bf r}_2$ in eq. (\ref{Gamma4}).   The quadratic part of the structure function makes no contribution to $Z$ and thus to the intensity fluctuations.  Physically this is because the quadratic term is associated with (linear) phase gradients which possess no curvature.  Since only the next most important term contributes to intensity fluctuations, this regime only applies to scattering in which the mass density variance is exceptionally high or the Fresnel scale is large.  Although it is demonstrated in the next section that strong scintillation is unlikely to occur even under the most extreme circumstances, the behaviour of the intensity fluctuation is still of formal interest.  The analysis is divided into two domains: (i) $0 < \beta < 2$, where the part of the phase structure function that contributes to intensity fluctuations scales as $r^{2+\beta}$, and is a factor $(r/L_0)^\beta$ smaller than the quadratic term, and (ii) $2 < \beta < 4$ where the dominant term scales as $r^4$ and is a factor of $(r/L_0)^2$ smaller than the quadratic term.

\subsubsection{$\beta < 2$}
A number of authors have investigated this case in the context of scintillation (Jakeman \& Jefferson 1984, Rino \& Owen 1984, Goodman \& Narayan 1985), so we merely summarize these results and recast them in the terms of the notation employed here.  
In the regime $r_0 \ll r_{\rm F}$ applicable to the strong perturbation case the intensity variance is independent of the scattering strength and is asymptotically equal to 
\begin{eqnarray}
\langle \Delta I^2 \rangle &=& \langle \Delta I_{\rm diff}^2 \rangle + \langle \Delta I_{\rm ref}^2 \rangle,  \\
&\null& \langle \Delta I_{\rm diff}^2 \rangle = \frac{2 \sqrt{\beta+1}}{2-\beta}, \qquad 0.2 \la \beta \la 1.8, \\
&\null& \langle \Delta I_{\rm ref}^2 \rangle = \frac{2 \sqrt{\beta+1}}{2-\beta} - 1,\qquad 0.2 \la \beta \la 1.8, \
\end{eqnarray}
where the subscripts `diff' and `ref' denote the character of the intensity variations as diffractive and refractive respectively.  In practice these results are valid for $r_{\rm F} / r_0 \ga 5$ (Jakeman \& Jefferson 1984), but the asymptotic solution is unreliable for values of $\beta$ near both 0 and 2.  

Refractive variations occur on the timescale,
\begin{eqnarray}
\tau_{\rm ref} \sim  r_{\rm F} v^{-1} \left[ \frac{(2+\beta) \pi}{2} \left( \frac{r_{\rm F}}{r_0} \right)^{2+\beta} \right]^{1/(2-\beta)} ,
\end{eqnarray}
while diffractive variations occur on the much shorter timescale, 
\begin{eqnarray}
\tau_{\rm diff} \sim r_{\rm F} \, v^{-1}  \left[ \frac{1}{2 (\beta+1)(\beta+2)} \left( \frac{r_0}{r_{\rm F}} \right)^{2+\beta} \right]^{1/(2-\beta)}.
\end{eqnarray}
Refractive variations are caused by phase curvature over the entire image of the scattered source as projected on the lensing plane.  There is a large amount of power on large scales because the phase power spectrum scales sharply as $q^{-4-\beta}$.  Following the argument advanced by Goodman \& Narayan (1985), the root mean square phase curvature per logarithmic interval scales as $\sim Q_0 K^2 q^{2-\beta} d(\log q)$, and structures on length scales $q$ focus at a distance $z \sim k/[ Q_0 K^2 q^{2-\beta} d(\log q)]^{1/2}$.  Thus the largest fluctuations which focus on the observer's plane, which are also most important due to the steep phase spectrum, are those at spatial wavenumbers $q_{\rm ref} \sim r_{\rm F}^{4/(\beta-2)} r_0^{(2+\beta)(2-\beta)}$.  This gives rise to lensing on timescales $\tau_{\rm ref} \sim v^{-1}  r_{\rm F}^{4/(2-\beta)} r_0^{(2+\beta)/(\beta-2)}$. 

Diffractive variations are caused by interference between radiation arriving from different parts of the lensed image.  This provides a physical argument for the timescale of diffractive changes.  The interference pattern changes on a timescale over which the relative phase delay between two patches separated a distance $\sim q_{\rm ref}^{-1}$ on the lensing plane changes by order one radian.  If both patches move a distance $\Delta x$ across the lensing plane the phase delay changes by $\sim [(q_{\rm ref}^{-1}+\Delta x)^2-(\Delta x)^2]/2 r_{\rm F}^2$.  Thus variations occur on a timescale $\tau \sim r_{\rm F}^2 q_{\rm ref}/ v$, which explains the magnitude of $\tau_{\rm diff}$ above.


The variance of the mutual coherence is computed by Goodman \& Narayan (1989) [cf. their eqs. (4.2.2) \& (2.5.12)] who find,
\begin{eqnarray}
\Gamma({\bf r},0) = \langle V({\bf r})^2 \rangle =  K +  
\frac{2^{1-\beta} \Gamma \left( 1- \frac{\beta}{2} \right)}{ (\beta+2) \sqrt{\beta+1} \Gamma \left( \frac{\beta}{2} \right) }  
\left( \frac{r}{r_0} \right)^2 \left( \frac{r \,r_0}{r_{\rm F}^2} \right)^{\beta}, 
\qquad  r \gg \left( \frac{r_0}{r_{\rm F}^2} \right)^{\beta/(2-\beta)}\, r_0^{2/(2-\beta)},
\end{eqnarray}
where $K$ is a constant that is not of interest here.  The variance in the mutual coherence decays over a much larger scale than the mean visibility itself.  The mean visibility decays exponentially quickly on a scale length $r_{\rm diff}$.  Recall that, unlike for intensity or visibility scintillations, the quadratic term in the structure function makes the dominant contribution in determining the ensemble average visibility (i.e. $\langle V({\bf r}) \rangle = \exp[-0.5 (r/r_{\rm diff})^2]$).  However, the mean square visibility varies on a much larger scale set by $r_0 \gg r_{\rm diff}$.  Simulations of scattering from such steep spectra (Narayan \& Goodman 1989) attribute this difference to the fact that the instantaneous lensed image has a fractal patchiness, meaning that an instantaneous image of the source appears fragmented into a number of highly-modulated subimages.  The ensemble-average image is necessarily smooth because it averages out all the effects of the large (linear) phase gradients and thus image wander, which are the primary cause of this fragmentation.


\subsubsection{$2 < \beta < 4$}

For yet steeper mass density power spectra the part of the phase structure function that contributes to intensity fluctuations scales as $r^4$, and the dominant contribution to the fourth-order moment of the wavefield reduces to
\begin{eqnarray}
\Gamma_4({\bf r}_1,{\bf r}_2) &=& \frac{1}{(2 \pi r_{\rm F}^2)^2} \int d^2{\bf r}_1' d^2{\bf r}_2' \exp 
\left[ \frac{i ({\bf r}_1'-{\bf r}_1) \cdot ({\bf r}_2-{\bf r}_2')}{r_{\rm F}^2} 
- 6 A ({\bf r}_1' \cdot {\bf r}_2')^2 - 2 A ({\bf r}_1' \times {\bf r}_2')^2 \right], \qquad A=\frac{4 \pi Q_0 K^2}{64 (\beta-2)}   q_{\rm min}^{2-\beta} \nonumber \\ \label{Gam4}.
 \end{eqnarray}
The variance of the intensity fluctuations is obtained by setting ${\bf r}_1={\bf r}_2 =0$ in eq. (\ref{Gam4}), and making the change of variables, $\Delta \theta= \theta_{x}-\theta_{y}$, $\Theta=(\theta_{x}+\theta_{y})/2$, $x = r_1'/r_{\rm F}$ and $y=r_2'/r_{\rm F}$, to obtain
\begin{eqnarray}
\langle \Delta I^2 \rangle + 1 = \frac{1}{2 \pi} \int_0^\infty dx dy \, x y \int_0^{2 \pi} d\Delta \theta \, \exp 
\left[ i x  y \cos \Delta \theta   
- 2 A r_{\rm F}^4 x^2 y^2 (1 + 2 \cos^2 \Delta \theta) \right].
\end{eqnarray}
Thus the amplitude of the intensity variations is controlled only by the dimensionless parameter $A r_{\rm F}^4$.  

The power spectrum of intensity fluctuations across the observer's plane is,
\begin{eqnarray}
W_I({\bf q}) = \int d^2{\bf r}_1' \exp \left\{- i {\bf r}_1' \cdot {\bf q} - 2 A r_{\rm F}^4 r_1'^2 q^2 [1+\cos^2( \theta_{r_1'} - \theta_q) ] \right\},
\end{eqnarray}
which is once again determined only by the dimensionless parameter $A r_{\rm F}^4$.  

No attempt is made to further analyse the characteristics of the intensity fluctuations in the regime of strong phase perturbations as this regime is unlikely to be encountered in the lensing of gravitational radiation even under the most extreme circumstances.


\section{Calculation of the temporal intensity fluctuation spectrum for lensing by stars} \label{Wcalc}

The contribution to the temporal intensity fluctuation spectrum for $\omega/v \la r_{\rm F}^{-1}$ from clustered stars is,
\begin{eqnarray}
W_{\rm coh}(\omega) 
&=& 
\frac{-2^{2-\gamma} \gamma  \Gamma \left(-\frac{\gamma}{2} \right) \pi^2 \sigma^2 M^2 K^2 r_{\rm F}^4 r_0^\gamma}{\Gamma \left( \frac{\gamma}{2} \right)\, v}
\int_0^{r_{\rm F}^{-1}} dq \, \left( \frac{\omega^2}{v^2} + q^2\right)^{\gamma/2-1} \left[ 1 - \left(\frac{\omega^2}{v^2} + q^2 \right) \frac{R^2}{5}\right] \nonumber \\
&=& \frac{2^{2-\gamma} \gamma  \Gamma \left(\frac{-\gamma}{2} \right) \pi^2 \sigma^2 M^2 K^2 r_{\rm F} r_0^\gamma}{15 \omega^2 v \, \Gamma \left( \frac{\gamma}{2} \right)} 
\left(\frac{\omega}{v} \right)^\gamma
\left[  
R^2 v^2 \null_2F_1 \left(\frac{3}{2},1-\frac{\gamma}{2}; \frac{5}{2}; \frac{-v^2}{r_{\rm F}^2 \omega^2} \right)
- 3 r_{\rm F}^2 (5 v^2 - R^2 \omega^2 ) 
\null_2F_1 \left(\frac{1}{2},1-\frac{\gamma}{2}; \frac{3}{2}; \frac{-v^2}{r_{\rm F}^2 \omega^2} \right) 
\right],  \nonumber \\
\end{eqnarray}
where we have written the stellar density profile to second-order in $q$, $|\tilde f_M({\bf q})|^2 \approx M^2 (1- q^2 R^2/5)$.
The power spectrum is simplified by recognising that the last argument of the two hypergeometric functions is large for $\omega < v r_{\rm F}^{-1}$.  Using the asymptotic expansion,
\begin{eqnarray}
\null_2F_1 (a,b;c;z) = \frac{\Gamma(b-a) \, \Gamma(c)}{\Gamma(b)\, \Gamma(c-a)} (-z)^{-a},
\end{eqnarray} 
keeping the leading order term in $\omega/v$, and neglecting terms of order $R^2/r_{\rm F}^2 \ll 1$ we obtain,
\begin{eqnarray}
W_{\rm coh} (\omega) = - 4 \gamma \, \Gamma(-\gamma) \,\pi^2\, \sin \left( \frac{\pi \gamma}{2} \right) \sigma^2 M^2 K^2 r_0^\gamma r_{\rm F}^4 \frac{1}{\omega} \left( \frac{\omega}{v} \right)^\gamma
,\qquad \frac{\omega}{v} < r_{\rm F}^{-1},\quad 1/2 < \gamma < 2. .
\end{eqnarray}
For higher angular frequencies $\omega > v r_{\rm F}^{-1}$, the temporal power spectrum is
\begin{eqnarray}
W_{\rm coh}(\omega) = \frac{2^{3-\gamma} \gamma \pi^2 K^2 M^2 \sigma^2 r_0^\gamma \Gamma \left(- \frac{\gamma}{2} \right)}{10 \omega^6 \Gamma \left( \frac{\gamma}{2} \right) }
v^3 \left( \frac{\omega}{v} \right)^{1+\gamma} e^{i \pi (1+\gamma)/2}
\left[ R^2 \omega^2 \beta_{\frac{-r_{\rm F}^2 \omega^2}{v^2}} \left( \frac{3-\gamma}{2}, \frac{\gamma}{2}-2\right)
+ ( 5v^2 - R^2 \omega^2 ) \beta_{\frac{-r_{\rm F}^2 \omega^2}{v^2}} \left( \frac{5-\gamma}{2}, \frac{\gamma}{2}-2\right)
 \right], \nonumber \\
\end{eqnarray}
where, upon expanding the incomplete beta function for large $z$,
\begin{eqnarray}
\beta_z(a,b) \approx \frac{\Gamma(a) \Gamma(1-a-b)}{\Gamma(1-b)} z^a (-z)^{-a} + \frac{1}{a+b-1} z^a (-z)^{b-1}.
\end{eqnarray}
neglecting terms ${\cal O}(R^2/r_{\rm F}^2)$ and keeping terms to lowest order in $\omega/v$ only one has
\begin{eqnarray}
W_{\rm coh}(\omega)  =  \frac{8\,\gamma (\gamma-3) (\gamma-1)}{8-6\gamma+\gamma^2}  
\Gamma(-\gamma) \sin \left( \frac{\pi \gamma}{2} \right) \pi^2 \sigma^2 K^2 M^2 r_0^\gamma \frac{1}{v} \left( \frac{\omega}{v} \right)^{\gamma-5}, \qquad \frac{\omega}{v} > r_{\rm F}^{-1}. 
\end{eqnarray}

\end{document}